\documentclass[11pt]{article}

\usepackage{graphicx}
\usepackage[footnotesize]{caption}

\usepackage{multirow}
\usepackage{dsfont}
\usepackage{amssymb}
\usepackage{slashed}
\usepackage{mathtools}
\usepackage{nicefrac}
\usepackage{verbatim}

\usepackage{color}

\def\beq{\begin{equation}}
\def\eeq{\end{equation}}
\newcommand{\be}{\begin{eqnarray}}

\newcommand{\ee}{\end{eqnarray}}
\newcommand{\ba}{\begin{array}}
\newcommand{\ea}{\end{array}}

\begin{document}
\begin{titlepage}

\vspace*{0.7cm}

\begin{center}
{
\bf\LARGE A Review of the Exceptional Supersymmetric \\[0.2cm]
Standard Model
}
\\[12mm]
Stephen F. King$^{1}$,
Stefano Moretti$^{1,2}$,
Roman Nevzorov $^{3,}$\footnote{nevzorov@itep.ru}
\\[-2mm]

\end{center}
\vspace*{0.50cm}
\centerline{$^{1}$ \it
School of Physics and Astronomy, University of Southampton, Highfield,}
\centerline{\it
Southampton, SO17 1BJ, United Kingdom }
\vspace*{0.2cm}
\centerline{$^{2}$\it
Particle Physics Department, Rutherford Appleton Laboratory, Chilton,}
\centerline{\it
Didcot, Oxon OX11 0QX, United Kingdom }
\vspace*{0.2cm}
\centerline{$^{3}$\it
NRC Kurchatov Institute --- ITEP, Moscow 117218, Russia}
\vspace*{1.20cm}

\begin{abstract}
\noindent
Local supersymmetry (SUSY) provides an attractive framework for the incorporation of
gravity and unification of gauge interactions within Grand Unified Theories (GUTs).
Its breakdown can lead to a variety of models with softly broken SUSY at low energies.
In this review article we focus on the SUSY extension of the Standard Model (SM) with an
extra $U(1)_{N}$ gauge symmetry originating from a string-inspired $E_6$ grand unified
gauge group. Only in this $U(1)$ extension of the minimal supersymmetric standard model (MSSM)
inspired by $E_6$ GUTs the right--handed neutrinos can be superheavy providing a mechanism
for the generation of the  lepton and baryon asymmetry of the Universe. The particle content of
this exceptional supersymmetric standard model (E$_6$SSM) includes three $27$ representations
of the $E_6$ group, to ensure anomaly cancellation, plus a pair of $SU(2)_W$ doublets
as required for gauge coupling unification. Thus E$_6$SSM involves extra exotic matter beyond
the MSSM. We consider symmetries that permit to suppress non-diagonal flavour transitions and
rapid proton decay, as well as gauge coupling unification, the breakdown of the gauge symmetry and the spectrum of
Higgs bosons in this model. The possible Large Hadron Collider (LHC) signatures caused by the presence of exotic states
are also discussed.
\end{abstract}

\end{titlepage}

\thispagestyle{empty}
\vfill
\newpage

\section{Introduction}

Symmetries play a key role in modern high energy physics. Indeed, it was realised a long time ago
that light hadron resonances form representations of the $SU(3)$ group, which is associated
with light quark flavours, while the physics of strong interactions is described by the coloured
$SU(3)_C$ gauge symmetry. It was also established that weak and electromagnetic forces represent
electroweak (EW) interactions based on the $SU(2)_W\times U(1)_Y$ gauge group. Within the
standard model (SM) of elementary particles, that describes perfectly the major part of all
experimental data measured in earth based experiments, $SU(2)_W\times U(1)_Y$ is spontaneously
broken to the abelian $U(1)_{em}$ gauge group associated with electromagnetism by means of the
Higgs mechanism. The latter predicts the existence of the Higgs boson which was recently discovered
at the LHC. Thus the Lagrangian of the SM is invariant under the Pointcar\'e group and
$SU(3)_C\times SU(2)_W\times U(1)_Y$ gauge symmetry transformations. The Pointcar\'e group is
an extension of Lorentz group that includes time and space translations whereas the transformations
of Lorentz group involve rotations about three axes and Lorentz boosts along them.

At very high energies the SM can be embedded into GUTs \cite{Georgi:1974sy}
based on the $SU(5)$ or $SO(10)$ gauge groups. In the case of  $SU(5)$ GUTs each SM family of quarks
and leptons fills in a complete one antifundamental and one antisymmetric second--rank tensor representations
of $SU(5)$, i.e. $\overline{5}+10$. Within  $SO(10)$ GUTs each family of  SM fermions
may belong to a single 16 dimensional spinor representation of $SO(10)$. Such models predict the existence
of right--handed neutrinos, allowing these to be used for both the see--saw mechanism \cite{Minkowski:1977sc}
and leptogenesis \cite{Fukugita:1986hr}.

SUSY GUTs permit to place fermions and bosons of the SM within one supermultiplet.
In order to achieve the unification of gauge interactions with gravity one needs to combine Pointcar\'e
and internal (gauge) symmetries. At the same time according to the Coleman-Mandula theorem the most
general symmetry which quantum field theory can have is a tensor product of the Pointcar\'e group
and an internal group \cite{Coleman:1967ad}. The Coleman-Mandula theorem can be overcome within
graded Lie algebras that have the following structure
\begin{equation}
\left[\hat{B},\hat{B}\right]=\hat{B},\qquad \left[\hat{B},\hat{F}\right]=\hat{F},\qquad
\left\{\hat{F},\hat{F}\right\}=\hat{B},
\label{1}
\end{equation}
where $\hat{B}$ and $\hat{F}$ are bosonic and fermionic generators. Graded Lie algebras that contain
the Pointcar\'e algebra are called supersymmetries. The simplest $N=1$ supersymmetry
involves a set of generators of the Pointcar\'e group (bosonic generators) and a single Weyl spinor operator
$Q_{\alpha}$ as well as its complex conjugate $Q_{\alpha}^{\dagger}=\overline{Q}_{\,\Dot{\alpha}}$ (fermionic
generators). SUSY algebra implies that each supermultiplet has the same number of bosonic and fermionic
degrees of freedom.

In $N=1$ SUSY GUTs based on the $E_6$ gauge group the complete fundamental
$27$ representation, which decomposes under $SO(10) \times U(1)_{\psi}$ subgroup as
\begin{equation}
27 \to \left(16,\,\frac{1}{\sqrt{24}}\right) \oplus
\left(10,\,-\frac{2}{\sqrt{24}}\right) \oplus
\left(1,\,\frac{4}{\sqrt{24}}\right)\,,
\label{2}
\end{equation}
contains one family of the SM fermions and Higgs doublet. The Higgs doublet is assigned to $\left(10,\,-\dfrac{2}{\sqrt{24}}\right)$.
The SM gauge bosons belong to the adjoint representation of $E_6$, i.e. a $78$--plet. In $N=2$ SUSY GUTs based on the $E_8$ gauge
group all SM particles belong to a $248$ dimensional representation of $E_8$ that decomposes under its $E_6$ subgroup as follows
\begin{equation}
248 \to 78 \oplus\, 3\times 27\, \oplus\, 3\times \overline{27}\, \oplus \, 8\times 1\,.
\label{3}
\end{equation}

The local SUSY (supergravity) leads to a partial unification of gauge interactions with gravity
\cite{susy-grav1,susy-grav2,susy-grav3}. However supergravity (SUGRA) is a non-renormalizable theory and has to be considered
as an effective low energy limit of some renormalizable or even finite theory. Currently, the best candidate for such
an underlying theory, i.e. the hypothetical single framework that explains and links together all physical aspects of the universe,
is a ten-dimensional heterotic superstring theory based on $E_8\times E'_8$ \cite{Green:1987sp}. Compactification of the
extra dimensions in this theory leads to an effective supergravity and results in the breakdown of $E_8$ to $E_6$
or its subgroups in the observable sector \cite{delAguila:1985hkb}. The remaining $E'_8$ plays the role of a hidden sector which
gives rise to spontaneous breakdown of SUGRA. As a consequence, a set of soft SUSY breaking terms
\cite{Barbieri:1982eh,Nilles:1982dy,Hall:1983iz,Soni:1983rm} characterized by the gravitino mass ($m_{3/2}$) is generated.
A large mass hierarchy between $m_{3/2}$ and the Planck scale $M_P$ can be caused by the non-perturbative effects in the hidden sector
that trigger the breakdown of local SUSY \cite{Nilles:1990zd}.

When $m_{3/2} \ll M_P$ the breakdown of the $E_6$ gauge group near the GUT scale $M_X$ may lead to a variety of SUSY models
at low energies including models based on the SM gauge group, like the MSSM, as well as extensions of the MSSM with an
extra $U(1)'$ gauge symmetry which is a linear combination of $U(1)_{\chi}$ and $U(1)_{\psi}$, i.e.:
\begin{equation}
U(1)'=U(1)_{\chi}\cos\theta_{E_6} + U(1)_{\psi}\sin\theta_{E_6}\,.
\label{4}
\end{equation}
Here $U(1)_\psi$ and $U(1)_\chi$ are associated with the subgroups
$E_6 \supset SO(10) \times U(1)_\psi \supset SU(5) \times U(1)_\chi \times U(1)_\psi$
whereas the SM gauge group is a subgroup of $SU(5)$, i.e. $SU(5) \supset SU(3)_C\times SU(2)_W\times U(1)_Y$.
In the simplest case $U(1)_{\chi}\times U(1)_{\psi}$ is broken down to its discrete subgroup
$Z_{2}^{M}=(-1)^{3(B-L)}$ which is the so--called matter parity. If in this case the low energy matter
content involves three families of the SM fermions and their scalar superpartners as well as
two $SU(2)_W$ doublets of the Higgs bosons ($H_1$ and $H_2$) and their fermionic partners (Higgsinos)
then this model corresponds to the simplest SUSY extension of the SM --- the MSSM.
Matter parity conservation implies that the lightest SUSY particle (LSP) is stable and can play the role
of dark matter. In order to reproduce the Higgs--fermion Yukawa interactions that induce the masses of all quarks
and charged leptons in the SM the MSSM superpotential has to include the following sum of the products of chiral superfields
\begin{equation}
W_{\rm MSSM} = y^U_{ab} Q_a u^c_b H_2 + y^D_{ab} Q_a d^c_b H_1 +
y^L_{ab} L_a e^c_b H_1 + \mu H_1 H_2\,,
\label{5}
\end{equation}
where $a$ and $b$ are family indices that run from 1 to 3. In Eq.~(\ref{5}) $Q_a$ and $L_a$ contain
the doublets of left--handed quark and lepton superfields, $e^c_a$, $u^c_a$ and $d^c_a$ are associated with
the right--handed lepton, up- and down--type quark superfields, respectively, while the Yukawa couplings
$y^U_{ab}$, $y^D_{ab}$ and $y^L_{ab}$ are dimensionless $3\times 3$ matrices in family space.
It was found that the EW and strong gauge couplings extracted from LEP data and extrapolated
to high energies using the renormalisation group (RG) equations do not meet within the SM but converge
to a common value near the scale $M_X\simeq 2\cdot 10^{16}\,\mbox{GeV}$ in the framework of the MSSM
\cite{Ellis:1990wk,Langacker:1991an,Amaldi:1991cn,Anselmo:1991uu}. This allows one to embed the MSSM into SUSY GUTs.

The MSSM superpotential in Eq.~(\ref{5}) contains only one bilinear term $\mu H_1 H_2$ which can be present
before SUSY is broken. Therefore one would naturally expect the parameter $\mu$ to be either zero or of the order
of the GUT scale $M_{X}$. If $\mu\sim M_{X}$ then the Higgs scalars get a huge positive contribution $\sim\mu^2$
to their squared masses and EW symmetry breaking (EWSB) does not occur. In contrast, when $\mu=0$ at some scale $Q$,
the mixing between Higgs doublets is not generated at any scale below $Q$ due to the non--renormalisation theorems
\cite{Salam:1974jj,Grisaru:1979wc}. In this case the minimum of the Higgs boson potential is attained for $<H_d>=0$
and the down--type quarks as well as the charged leptons remain massless. In order to ensure the correct pattern
of the EW symmetry breaking (EWSB), $\mu$ is required to be of the order of the SUSY breaking scale $M_S$.

In the framework of the simplest extension of the MSSM, the next--to--MSSM (NMSSM),
the superpotential is invariant with respect to the discrete transformations $\Phi_{i}\to e^{2\pi i/3}\Phi_{i}$
of the $Z_3$ group. The term $\mu (H_1 H_2)$ does not meet this requirement and therefore can not be included.
The superpotential of the NMSSM is given by \cite{Ellwanger:2009dp}
\begin{equation}
W_{\rm NMSSM}= \lambda S( H_1 H_2 ) + \dfrac{\kappa}{3} S^3 + W_{\rm MSSM}(\mu=0),
\label{6}
\end{equation}
where $S$ is an extra singlet superfield. It acquires a vacuum expectation value (VEV), i.e. $\langle S \rangle=s/\sqrt{2}$, and an effective
$\mu$ parameter is generated ($\mu=\lambda s/\sqrt{2}\sim M_S$). The cubic term of the new singlet superfield $S$ in
the superpotential (\ref{6}) explicitly breaks an additional global $U(1)$ symmetry which is a common way to avoid the appearance of
the axion in the particle spectrum. However the NMSSM itself is not without problems. The VEVs of the Higgs fields break the exact $Z_3$
symmetry leading to the formation of domain walls in the early universe between regions which were causally disconnected during the period
of EWSB \cite{Zeldovich:1974uw}. Such domain structure of vacuum creates unacceptably large anisotropies in the cosmic microwave background
radiation \cite{Vilenkin:1984ib}. Because of this the NMSSM superpotential should contain additional operators that violate the $Z_3$ symmetry
and prevent the appearance of domain walls \cite{Panagiotakopoulos:1998yw,Panagiotakopoulos:1999ah}.

In the $U(1)'$ extensions of the MSSM inspired by $E_6$ the extra $U(1)'$ gauge symmetry (\ref{4}) forbids
an elementary $\mu$ term if $\theta_{E_6} \ne 0$ or $\pi$. Nevertheless these extensions of the SM allow the interaction
$\lambda S(H_d H_u)$ in the superpotential while the $S^3$ term is forbidden by the $U(1)'$ gauge symmetry.
Near the scale $M_S$ the scalar component of the SM singlet superfield $S$ develops a non--zero VEV breaking $U(1)'$
and an effective $\mu$ term of the required size is automatically generated. Clearly there are no domain wall problems
in such models since there is no discrete $Z_3$ symmetry. Different aspects of the phenomenology of  $E_6$ inspired SUSY
models have been extensively studied in the past \cite{Hewett:1988xc,01,02,03,04,05,06,07,08,09,010}.
Previously, the implications of $E_6$ inspired SUSY models with an additional $U(1)'$ gauge symmetry have been studied
for the EWSB \cite{Langacker:1998tc,Cvetic:1995rj,Cvetic:1996mf,Cvetic:1997ky,Suematsu:1994qm,Keith:1997zb,Daikoku:2000ep},
neutrino physics \cite{Kang:2004ix, Ma:1995xk}, fermion mass hierarchy and mixing \cite{Stech:2008wd}, leptogenesis \cite{Hambye:2000bn,King:2008qb},
EW baryogenesis \cite{Ma:2000jf,Kang:2004pp}, the $Z'$ mass limits \cite{Accomando:2010fz}, collider signatures associated
with the exotic quarks and squarks \cite{Kang:2007ib}, the muon anomalous magnetic moment \cite{g-2-1,g-2-2},
the electric dipole moment of the electron \cite{Suematsu:1997tv} and of the tau lepton \cite{GutierrezRodriguez:2006hb},
 lepton flavor violating processes like $\mu\to e\gamma$ \cite{Suematsu:1997qt} and  CP-violation in the
Higgs sector \cite{Ham:2008fx}. The neutralino sector in $E_6$ inspired SUSY models was examined in
\cite{Keith:1997zb,Suematsu:1997tv,GutierrezRodriguez:2006hb,Suematsu:1997qt,Suematsu:1997au,Keith:1996fv,n1,n2,n3,n4,Gherghetta:1996yr,E6neutralino-higgs}.
The Higgs sector and the theoretical upper bound on the lightest Higgs boson mass in the $E_6$ inspired SUSY models were explored
in \cite{Daikoku:2000ep,E6neutralino-higgs, King:2005jy,King:2005my,E6-higgs}.

In this review article we consider a specific $E_6$ inspired SUSY realisation of the above $U(1)'$ type model associated
with $\theta_{E_6}=\arctan\sqrt{15}$. This choice of Abelian $U(1)'$ corresponds to $U(1)_{N}$ gauge symmetry. Thus such a
SUSY model is based on the SM gauge group together with an additional $U(1)_{N}$ factor. In this exceptional supersymmetric
standard model (E$_6$SSM) \cite{King:2005jy,King:2005my} right-handed neutrinos do not participate in the gauge interactions.
Therefore only in such a $U(1)'$ extension of the MSSM inspired by $E_6$ GUTs the right--handed neutrinos can be superheavy,
so that a see-saw mechanism can be used to generate the mass hierarchy in the lepton sector, providing a comprehensive
understanding of the neutrino oscillations data. Successful leptogenesis is also a distinctive feature of the E$_6$SSM since
the heavy Majorana right-handed neutrinos may decay into final states with lepton number $L=\pm 1$, creating a lepton asymmetry
in the early Universe \cite{Hambye:2000bn,King:2008qb,Nevzorov:2017gir}.

The layout of this paper is as follows. In the next Section we specify the $U(1)_{N}$ extensions of the MSSM and discuss global
symmetries that prevent non-diagonal flavour transitions as well as rapid proton decay in these SUSY models. The two--loop
RG flow of the gauge couplings within the E$_6$SSM is examined in Section 3. The Higgs sector dynamics and the emerging spectrum of the masses and  couplings of the Higgs bosons
are discussed in Sections 4 and 5, respectively. In section 6 the possible LHC signatures of the E$_6$SSM are considered.
Section 7 is reserved for our conclusions and outlook.

\section{The $U(1)_{N}$ extensions of the MSSM}

The~ E$_6$SSM~ implies~ that~ near~ the~ GUT~ scale~ $E_6$~ or~ its~ subgroup~ is~ broken~ down~ to~\\
$SU(3)_C\times SU(2)_W\times U(1)_Y\times U(1)_{N}\times Z_{2}^{M}$ \cite{King:2005jy,King:2005my}.
With additional Abelian gauge symmetries it is important to ensure the cancellation of anomalies.
In any model based on the subgroup of $E_6$ the anomalies are canceled automatically if the low-energy
spectrum involves complete representations of $E_6$. Consequently, in the E$_6$SSM the particle
spectrum is extended by a number of exotics which, together with ordinary quarks and leptons, form
three complete 27-dimensional representations of $E_6$, referred to as $27_i$ with $i=1,2,3$.
These multiplets decompose under the $SU(5)\times U(1)_{N}$ subgroup of $E_6$ as follows:
\begin{equation}
\begin{array}{c}
27_i\to \left(10,\,\dfrac{1}{\sqrt{40}}\right)_i + \left(5^{*},\,\dfrac{2}{\sqrt{40}}\right)_i
+ \left(5^{*},\,-\dfrac{3}{\sqrt{40}}\right)_i + \left(5,-\dfrac{2}{\sqrt{40}}\right)_i
+\left(1,\dfrac{5}{\sqrt{40}}\right)_i+\left(1,0\right)_i\,.
\end{array}
\label{7}
\end{equation}
The first and second quantities in the brackets are the $SU(5)$ representation and extra $U(1)_{N}$ charge respectively.
An ordinary SM family, which contains the doublets of left-handed quarks $Q_i$ and leptons $L_i$, right-handed
up- and down-quarks ($u^c_i$ and $d^c_i$) as well as right-handed charged leptons $(e^c_i)$, is assigned to $\left(10,\dfrac{1}{\sqrt{40}}\right)_i
+\left(5^{*},\,\dfrac{2}{\sqrt{40}}\right)_i$. Right-handed neutrinos $N^c_i$ are associated with the last term in Eq.~(\ref{7}), $\left(1,0\right)_i$.
The next-to-last term, $\left(1,\frac{5}{\sqrt{40}}\right)_i$, represents new SM-singlet fields $S_i$, with non-zero $U(1)_{N}$ charges
that therefore survive down to the EW scale. The pair of $SU(2)_W$-doublets ($H^d_{i}$ and $H^u_{i}$) that are contained in $\left(5^{*},\,-\dfrac{3}{\sqrt{40}}\right)_i$ and $\left(5,-\dfrac{2}{\sqrt{40}}\right)_i$ have the quantum numbers of Higgs doublets.
They form either Higgs or Inert Higgs $SU(2)_W$ multiplets, i.e. Higgs--like doublets that do not develop VEVs.
Other components of these $SU(5)$ multiplets form colour triplets of exotic quarks $\overline{D}_i$ and $D_i$ with electric
charges $+ 1/3$ and $-1/3$ respectively. These exotic quark states carry a $B-L$ charge $\left(\pm\dfrac{2}{3}\right)$ twice larger
than that of ordinary ones. Therefore in phenomenologically viable $U(1)_{N}$ extensions of the MSSM they can be either diquarks or leptoquarks.

In addition to the complete $27_i$ multiplets the splitting of $27'_l$ and $\overline{27'}_l$ within the $E_6$ GUTs can give rise to a set
of $M_{l}$ and $\overline{M}_l$ supermultiplets with opposite quantum numbers at low energies. In the simplest case the low energy particle
spectrum of the E$_6$SSM is supplemented by $SU(2)_W$ doublet $L_4$ and anti-doublet $\overline{L}_4$ states from the extra $27'$ and
$\overline{27'}$ to preserve gauge coupling unification, where $L_4$ supermultiplet has the quantum numbers of left-handed leptons.
Thus, in addition to a $Z'$ corresponding to the $U(1)_N$ symmetry, the E$_6$SSM involves extra matter beyond the MSSM that fill in
three $5+5^{*}$ representations of $SU(5)$ plus three $SU(5)$ singlets with $U(1)_N$ charges. The gauge group and field content of the E$_6$SSM
can originate from the orbifold GUT models \cite{Nevzorov:2012hs,Nevzorov:2018leq}.

Over the last fifteen years, several variants of the E$_6$SSM have been proposed
\cite{King:2005jy,King:2005my,Howl:2007hq,Howl:2007zi,Howl:2008xz,Howl:2009ds, Athron:2010zz,Hall:2011zq,Callaghan:2012rv,
Nevzorov:2012hs,Callaghan:2013kaa,Athron:2014pua,King:2016wep}.
The $E_6$ inspired SUSY models with an additional $U(1)_{N}$ gauge symmetry have been thoroughly
investigated as well. They have been studied in \cite{Ma:1995xk} in the context of non-standard neutrino models,
in \cite{Suematsu:1997au} from the point of view of $Z-Z'$ mixing, in \cite{Keith:1997zb,Suematsu:1997au,Keith:1996fv}
the neutralino sector was explored, in \cite{Hall:2009aj} the implications of the exotic states for the dark matter
was considered, in \cite{Keith:1997zb,King:2007uj} the RG flow of the couplings was examined, and in
\cite{Suematsu:1994qm,Keith:1997zb,Daikoku:2000ep} EWSB was investigated. More recently, the RG flow of the
Yukawa couplings and the theoretical upper bound on the lightest Higgs boson mass were explored in the vicinity
of the quasi--fixed point \cite{Nevzorov:2013ixa,Nevzorov:2015iya} that appears as a result of the intersection of
the invariant and quasi--fixed lines \cite{Nevzorov:2001vj,Nevzorov:2002ub}. Detailed studies of the E$_6$SSM have established
that extra exotic matter and $Z^\prime$ predicted by this model may give rise to distinctive LHC signatures
\cite{King:2005jy,King:2005my,Howl:2007zi,Athron:2010zz,King:2006vu,King:2006rh,Athron:2011wu,Belyaev:2012si,Belyaev:2012jz},
as well as may lead to non-standard Higgs decays for sufficiently light exotics
\cite{Athron:2014pua,Nevzorov:2015iya,Nevzorov:2013tta,Hall:2010ix,Hall:2010ny,Hall:2011au,Hall:2013bua,Nevzorov:2014sha,Athron:2016usd}.
Within the constrained version of the E$_6$SSM (cE$_6$SSM) and its modifications the particle spectrum and associated phenomenological implications
were explored in \cite{Athron:2011wu,Athron:2008np,Athron:2009ue,Athron:2009bs,Athron:2012sq,Athron:2015vxg,Athron:2016gor}
while the degree of fine tuning was examined in \cite{Athron:2013ipa,Athron:2015tsa}. The threshold corrections to the running gauge
and Yukawa couplings in the E$_6$SSM and their impact in the cE$_6$SSM were studied in \cite{Athron:2012pw}.
The renormalisation of the VEVs in the E$_6$SSM was considered in \cite{Sperling:2013eva,Sperling:2013xqa}.

The superpotential of the $U(1)_{N}$ extensions of the MSSM contains the renormalisable part that comes from
the $27\times 27\times 27$ decomposition of the $E_6$ fundamental representation and can be written as
\begin{equation}
\begin{array}{rcl}
W_{E_6} & = & W_0+W_1+W_2\,,\\
W_0 & = & \lambda_{ijk} S_i (H^d_{j} H^u_{k})+\kappa_{ijk} S_i (D_j \overline{D}_k)+
h^N_{ijk} N_i^c (H^u_{j} L_k)+ h^U_{ijk} u^c_{i} (H^u_{j} Q_k) +\\[0mm]
& + & h^D_{ijk} d^c_i (H^d_{j} Q_k) +  h^E_{ijk} e^c_{i} (H^d_{j} L_k)\,,\\[0mm]
W_1&=& g^Q_{ijk} D_{i} (Q_j Q_k)+g^{q}_{ijk}\overline{D}_i d^c_j u^c_k\,,\\[0mm]
W_2&=& g^N_{ijk}N_i^c D_j d^c_k+g^E_{ijk} e^c_i
D_j u^c_k+g^D_{ijk} (Q_i L_j) \overline{D}_k\,.
\end{array}
\label{8}
\end{equation}
In Eq.~(\ref{8}) the summation over repeated family indexes ($i,j,k=1,2,3$) is implied. The part of the superpotential (\ref{8}) possesses
a global $U(1)$ symmetry which is associated with $B-L$ number conservation. This $U(1)$ symmetry is a linear superposition of $U(1)_Y$ and
$U(1)_{\chi}$. On the other hand if terms in $W_1$ and $W_2$ are simultaneously present in the superpotential then baryon and lepton numbers
are violated. In other words one cannot define the baryon and lepton numbers of the exotic quarks $D_i$ and $\overline{D}_i$ so that
the complete Lagrangian is invariant separately under $U(1)_{B}$ and $U(1)_{L}$ global symmetries. Thus as in any other SUSY extension of the
SM the gauge symmetry of the models under consideration does not forbid lepton and baryon number violating operators. Because of this
all these models in general suffer from problems related with rapid proton decay.

Moreover exotic states in the $U(1)_{N}$ extensions of the MSSM give rise to new Yukawa interactions that may induce unacceptably large
flavor changing processes. Indeed, in the most general case three families of $H^u_{i}$ and $H^d_{i}$ can couple to ordinary quarks and charged
leptons of different generations resulting in the phenomenologically unwanted non--diagonal flavor transitions even at the tree level. Such
non--diagonal flavor interactions contribute to the amplitude of $K^0-\overline{K}^0$ oscillations and give rise to new channels of muon decay like
$\mu\to e^{-}e^{+}e^{-}$. In order to suppress flavor changing neutral currents (FCNCs) one can postulate $Z^{H}_2$ symmetry.
If all matter supermultiplets except one pair of $H^u_{i}$ and $H^d_{i}$ (say $H_d\equiv H^d_{3}$ and $H_u\equiv H^u_{3}$) as well as
one SM--type singlet superfield ($S\equiv S_3$) are odd under this symmetry then only $H_u$ couples to up--type quarks and
only $H_d$ interacts with the down--type quarks and charged leptons \cite{King:2005jy,King:2005my}. The couplings of all other exotic states
to the ordinary quark and lepton supermultiplets are forbidden that eliminates any problems related with the non--diagonal flavour transitions
at the tree level. In this original E$_6$SSM model the scalar components of the supermultiplets $H_u$, $H_d$ and $S$ compose the Higgs
sector. In particular, the third family SM--singlet superfield $S_3$ gets a VEV, $\langle S_3 \rangle = s/\sqrt{2}$, breaking
$U(1)_{N}$ gauge symmetry. This VEV is responsible for the effective $\mu$ term and D-fermion masses. The first and second families of Higgs doublets
and SM-singlets, which do not get VEVs, are called ``inert''. At the same time the modified version of the E$_6$SSM, in which
three SM-singlet superfields $S_i$ are taken to be even under the $Z^{H}_2$ symmetry, was also recently considered \cite{King:2016wep}.
In this case all superfields $S_i$ develop VEVs. They couple to $H_u$, $H_d$ as well as other exotic bosons and fermions.

Although the $Z^{H}_2$ symmetry forbids not only flavor changing processes but also the most dangerous baryon and
lepton number violating operators, it can not be an exact symmetry. Indeed, this symmetry forbids all Yukawa interactions in $W_1$ and
$W_2$ that allow the lightest exotic quarks to decay. The Lagrangian of such model is invariant not only with respect to $U(1)_L$ and
$U(1)_B$ but also under $U(1)_D$ symmetry transformations
\begin{equation}
D\to e^{i\alpha} D\,,\qquad\qquad \overline{D}\to e^{-i\alpha}\overline{D}\,.
\label{9}
\end{equation}

The $U(1)_D$ invariance ensures that the lightest exotic quark is extremely long--lived. The $U(1)_L$, $U(1)_B$ and $U(1)_D$ global symmetries are
expected to be broken by the non--renormalizable operators which are suppressed by inverse power of the GUT scale $M_X$.
Since $E_6$ forbids any dimension five operators that break $U(1)_D$ global symmetry the lifetime of the lightest exotic quarks is expected to be
of order of
\begin{equation}
\tau_D > M_X^4/\mu_D^5\,,
\label{10}
\end{equation}
where $\mu_D$ is the mass of the lightest exotic quark. When $\mu_D\simeq \mbox{TeV}$ the lifetime of the lightest exotic quarks
$\tau_D > 10^{49}\,\mbox{GeV}^{-1}\sim 10^{17}\,\mbox{years}$, i.e. it is considerably larger than the age of the Universe.
So long--lived exotic quarks would have been copiously produced during the very early epochs of the Big Bang. Those lightest exotic
quarks which survive annihilation would have been confined in heavy hadrons which would annihilate further. The remaining heavy
hadrons with exotic quarks originating from the Big Bang should be present in terrestrial matter. Various theoretical
estimates \cite{Wolfram:1978,Dover:1979sn} show that if such remnant particles in the mass range from $1\,\mbox{GeV}$ to $10\,\mbox{TeV}$
would exist in nature, today their concentration is expected to be at the level of $10^{-10}$ per nucleon. At the same time different
experiments set stringent upper limits on the relative concentrations of such nuclear isotopes which vary from $10^{-15}$ to $10^{-30}$ per
nucleon \cite{Rich:1987jd,Smith:1988ni,Hemmick:1989ns}. Therefore the extensions of the SM with so long-lived exotic quarks are basically ruled out.
This means that the discrete $Z^{H}_2$ symmetry can only be an approximate one.

To prevent rapid proton decay within the $U(1)_{N}$ extensions of the MSSM one can impose
either $Z_2^L$ or $Z_2^B$ discrete symmetry. If the Lagrangian is invariant with respect to an exact $Z_2^L$ symmetry, under which all
superfields except lepton ones (including $L_4$ and $\overline{L}_4$) are even, then all Yukawa interactions in $W_2$ are forbidden and
the baryon number conservation requires the exotic quarks to be diquarks (Model I). In this case the most general renormalisable superpotential
which is allowed by the $SU(3)_C\times SU(2)_W\times U(1)_{Y}\times U(1)_{N}$ gauge symmetry can be presented in the following form:
\begin{equation}
\begin{array}{rcl}
W_{\rm{E}_6\rm{SSM\, I}}&=& W_0+W_1+\dfrac{1}{2}M_{ij}N_i^c N_j^c+W'_0\,,\\[2mm]
W'_0 & = & \mu_L L_4\overline{L}_4 + \tilde{h}^L_{ij} e^c_{i} (H^d_{j} L_4) + h^L_{ij} N_i^c (H^u_{j} L_4)\,.
\end{array}
\label{11}
\end{equation}
The terms in $W'_0$ are caused by the splitting of $27'$ and $\overline{27'}$ representations of $E_6$.
An alternative possibility is to assume that the exotic quarks $D_i$ and $\overline{D_i}$ as well as ordinary lepton superfields,
$L_4$ and $\overline{L}_4$ are all odd under $Z_2^B$ whereas the others remain even. As a consequence all terms in $W_1$ are ruled out by
the discrete $Z_2^B$ symmetry and exotic quarks carry baryon ($B_{D}=1/3$ and $B_{\overline{D}}=-1/3$) and lepton ($L_{D}=1$ and $L_{\overline{D}}=-1$) numbers simultaneously (Model II). Thus in Model II $D_i$ and $\overline{D_i}$ are leptoquarks. The most general renormalisable
superpotential in Model II are given by
\begin{equation}
W_{\rm{E}_6\rm{SSM\, II}}= W_0 + W_2 + \dfrac{1}{2}M_{ij}N_i^c N_j^c + W'_0 + g^L_{ik} (Q_i L_4) \overline{D}_k\,,
\label{12}
\end{equation}
The last term in Eq.~(\ref{12}) appears because of the splitting of $27'$.
In the superpotentials (\ref{11})-(\ref{12}) the $SU(2)_W$ doublet $L_4$ is redefined in such a way that
$W'_0$ contains only one bilinear term. The mass parameter $\mu_L$ should not be too large otherwise it spoils gauge coupling unification.
Within SUGRA models the appropriate term $\mu_L L_4\overline{L}_4$ in the superpotentials (\ref{11})-(\ref{12}) can be induced if the
K\"ahler potential contains an extra term $(Z L_4\overline{L}_4 + h.c)$ \cite{Giudice:1988yz,Casas:1992mk}. This is the same mechanism
which is used in the MSSM to solve the $\mu$ problem. Within the $U(1)_{N}$ extensions of the MSSM the bilinear term
involving $H_d$ and $H_u$ are forbidden by the $U(1)_{N}$ gauge symmetry so that the mechanism mentioned above cannot be applied for
the generation of $\mu H_d H_u$ in the E$_6$SSM superpotential.

The superpotentials of the Models I and II also include bilinear terms, $\dfrac{1}{2}M_{ij} N_i^c N_j^c$,
responsible for the right--handed neutrino masses. The corresponding mass parameters $M_{ij}$ are expected to be at intermediate
mass scales. They can be induced through the non--renormalisable interactions of the form
\begin{equation}
\delta W=\dfrac{\kappa_{ij}}{M_{Pl}}(\overline{27}_H\, 27_i)(\overline{27}_H\, 27_j)\qquad \Longrightarrow\qquad
M_{ij}=\dfrac{2\kappa_{ij}}{M_{Pl}}<\overline{N}_H^c>^2\,,
\label{13}
\end{equation}
if the $N^c_H$ and $\overline{N}_H^c$ components of some extra $27_H$ and $\overline{27}_H$ representations develop VEVs
along the $D$--flat direction $<N_H^c>=<\overline{N}_H^c>$. These VEVs can also break $U(1)_{\psi}\times U(1)_{\chi}$
down to $U(1)_{N}\times Z_{2}^{M}$ symmetry \cite{Nevzorov:2012hs}. To get a reasonable pattern for the left--handed neutrino
masses and mixing such breakdown should take place somewhere around the GUT scale $M_X$.

The superpotentials (\ref{11})-(\ref{12}) involve a lot of new Yukawa couplings in comparison to the SM.
In general the exact $Z_2^L$ and $Z_2^B$ discrete symmetries do not guarantee the absence of FCNCs in the $U(1)_{N}$ extensions
of the MSSM. At the same time it is worth noting that the observed mass hierarchy of quarks and charged leptons implies that most
of the Yukawa couplings in the SM and MSSM are small. Therefore it is natural to assume some hierarchical structure
of the Yukawa interactions that may permit to suppress non--diagonal flavor transitions. Also it seems reasonable to use
the approximate $Z^{H}_2$ symmetry to eliminate problems related with flavor changing processes. The appropriate suppression of
the non--diagonal flavor interactions can be achieved if all $Z^{H}_2$ symmetry violating couplings are less than $10^{-4}$.
In the limit when all Yukawa couplings, that explicitly break the $Z^{H}_2$ symmetry, are negligibly small, the superpotential
of the E$_6$SSM reduces to
\begin{equation}
\begin{array}{rcl}
W_{\rm{E}_6\rm{SSM}} & = &  \lambda S (H_u H_d) + \lambda_{\alpha\beta} S (H^d_{\alpha} H^u_{\beta})+
\kappa_{ij} S (D_{i} \overline{D}_{j}) + \tilde{f}_{\alpha\beta} S_{\alpha} (H^d_{\beta}H_u) + f_{\alpha\beta} S_{\alpha} (H_d H^u_{\beta})\\
& + & \mu_L L_4\overline{L}_4 + \dfrac{1}{2}M_{ij}N_i^c N_j^c + W_{L_4} + W_{\rm MSSM}(\mu=0)\,,\\
\end{array}
\label{14}
\end{equation}
where
\begin{equation}
W_{L_4}  =   \tilde{h}^L_{i3} e^c_{i} (H_d L_4) + h^L_{i3} N_i^c (H_u L_4)\,,
\label{15}
\end{equation}
$\alpha,\beta=1,2$ and $i,j=1,2,3$\,. If some of the couplings $\lambda$, $\lambda_{\alpha\beta}$ or
$\kappa_{ij}$ are rather large at the GUT scale $M_X$, they affect the evolution of the soft scalar mass $m_S^2$ of the
singlet field $S$ quite strongly resulting in negative values of $m_S^2$ at low energies. This triggers the breakdown of
$U(1)_{N}$ gauge symmetry. The singlet VEV must be large enough to generate sufficiently large masses of the $Z'$ boson
and exotic particles. This also implies that the Yukawa couplings $\lambda$, $\lambda_{\alpha\beta}$ and $\kappa_{ij}$
have to be large enough. On the other hand the large value of the top--quark Yukawa coupling provides a radiative mechanism
for generating the VEVs of $H_u$ and $H_d$ that break the $SU(2)_W\times U(1)_{Y}$ gauge symmetry.

Since in the $U(1)_{N}$ extensions of the MSSM the $Z_{2}^{M}$ symmetry and $R$--parity are conserved the lightest $R$--parity odd state, i.e.
the lightest SUSY particle (LSP), must be stable. Using the method proposed in \cite{Hesselbach:2007te,Hesselbach:2007ta,Hesselbach:2008vt} it
was shown that the LSP and next--to--lightest SUSY particle (NLSP) in the E$_6$SSM have masses below $60-65\,\mbox{GeV}$ \cite{Hall:2010ix}.
The LSP and NLSP ($\tilde{H}^0_1$ and $\tilde{H}^0_2$) are predominantly linear superpositions of the fermion components of the two SM
singlet superfields $S_{\alpha}$. Although the couplings of $\tilde{H}^0_1$ to the SM gauge bosons and fermions are quite small
LSP could account for all or some of the observed cold dark matter density if it had a mass close to half the $Z$ mass.
In this case LSP annihilate mainly through an $s$--channel $Z$--boson \cite{Hall:2010ix}. However the SM-like Higgs boson decays
more than 95\% of the time into either $\tilde{H}^0_1$ or $\tilde{H}^0_2$ in these scenarios while all other branching ratios
would be strongly suppressed. Nowadays such scenario are ruled out by the LHC experiments. If fermion components of the SM
singlet superfields $S_{\alpha}$ are substantially lighter than $M_Z$ the annihilation cross section for
$\tilde{H}^0_1\tilde{H}^0_1\to \mbox{SM particles}$ becomes too small leading to the cold dark matter density that is much larger
than its measured value.

Nevertheless in the E$_6$SSM with approximate $Z_2^H$ symmetry one of the lightest $R$--parity odd state can account for all or
some of the observed cold dark matter density. In order to prevent the decays of this state into the LSP and NLSP an additional
$Z^S_2$ symmetry needs to be postulated \cite{Hall:2011zq}. In the corresponding variant of the E$_6$SSM couplings $\tilde{f}_{\alpha\beta}$
and $f_{\alpha\beta}$ vanish. As a result the fermion components of the SM singlet superfields $S_{\alpha}$ remain massless and
decouple. If $Z'$ boson is sufficiently heavy the presence of these massless states does not affect Big Bang Nucleosynthesis (BBN) \cite{Hall:2011zq}.
Since $\tilde{f}_{\alpha\beta}=f_{\alpha\beta}=0$ the branching ratios of the SM--like Higgs decays into $\tilde{H}^0_1$ and $\tilde{H}^0_2$ vanish.

\begin{table}[ht]
\centering
\begin{tabular}{|c|c|c|c|c|c|c|c|}
\hline
& $Q_i, u^c_i, d^c_i$	& $L_i, e^c_i, N^c_i$ & $\overline{D}_i, D_i$ & $H^d_{\alpha}, H^u_{\alpha}$ & $ S_{\alpha} $ &
$H_d, H_u, S $     & $L_4, \overline{L}_4$\\
\hline
$Z^{H}_2$	& $-$ & $-$ & $-$ & $-$ & $-$ & $+$ & $-$\\
$Z_2^L$		& $+$ & $-$ & $+$ & $+$ & $+$ & $+$ & $-$\\
$Z_2^B$		& $+$ & $-$ & $-$ & $+$ & $+$ & $+$ & $-$\\
$Z_2^S$		& $+$ & $+$ & $+$ & $+$ & $-$ & $+$ & $+$\\
$\tilde{Z}^H_2$	& $-$ & $-$ & $-$ & $-$ & $-$ & $+$ & $+$\\
$Z^{M}_2$	& $-$ & $-$ & $+$ & $+$ & $+$ & $+$ & $-$\\
$Z_{2}^{E}$	& $+$ & $+$ & $-$ & $-$ & $-$ & $+$ & $-$\\
\hline
\end{tabular}
\caption{Transformation properties of matter supermultiplets
under $Z^{H}_2$, $Z_2^L$, $Z_2^B$, $Z^S_2$, $\tilde{Z}^H_2$, $Z_{2}^{M}$ and $Z_{2}^{E}$ discrete symmetries in the E$_6$SSM.}
\label{tab1}
\end{table}

Instead of $Z^{H}_2$, $Z_2^L$ and $Z_2^B$ one can impose a single discrete $\tilde{Z}^{H}_2$ symmetry which forbids
tree-level flavor-changing transitions and the most dangerous operators that violate baryon and lepton numbers.
In this case $H_u$, $H_d$, $S$, $L_4$ and $\overline{L}_4$ are even under the $\tilde{Z}^{H}_2$ symmetry
while all other supermultiplets are odd \cite{Nevzorov:2012hs}. Neglecting all suppressed non-renormalisable interactions, the
superpotential of this variant of the E$_6$SSM is given by Eq.~(\ref{14}) with
\begin{equation}
W_{L_4} = g^L_{ik} (Q_i L_4) \overline{D}_k + \tilde{h}^L_{i\alpha} e^c_{i} (H^d_{\alpha} L_4) + h^L_{i\alpha} N_i^c (H^u_{\alpha} L_4)\,,
\label{16}
\end{equation}
where $\alpha=1,2$ and $i,k=1,2,3$\,. Since the low--energy effective Lagrangian of this SUSY models is
invariant under both $Z_{2}^{M}$ and $\tilde{Z}^{H}_2$ symmetries and $\tilde{Z}^{H}_2 = Z_{2}^{M}\times Z_{2}^{E}$,
the $Z_{2}^{E}$ symmetry associated with exotic states is also conserved. The transformation properties of different
components of $27_i$ supermultiplets under the $Z^{H}_2$, $Z_2^L$, $Z_2^B$, $Z^S_2$, $\tilde{Z}^{H}_2$, $Z_{2}^{M}$ and $Z_{2}^{E}$
symmetries are summarized in Table~\ref{tab1}. The $Z_{2}^{E}$ symmetry conservation ensures that the lightest exotic state,
which is odd under this symmetry, is stable. The simplest phenomenologically viable scenarios imply that
$f_{\alpha\beta}\sim \tilde{f}_{\alpha\beta}< 10^{-6}$. As a consequence two lightest exotic states ($\tilde{H}^0_1$ and $\tilde{H}^0_2$),
which are formed by the fermion components of the superfields $S_{\alpha}$, are substantially lighter than $1\,\mbox{eV}$.
They compose hot dark matter in the Universe but gives only a very minor contribution to the dark matter density \cite{Nevzorov:2012hs}.
The presence of very light neutral fermions in the particle spectrum might also have interesting implications
for the neutrino physics (see, for example \cite{Frere:1996gb}). The invariance of the Lagrangian under
the $Z_{2}^{M}$ ensures that the lightest $R$--parity odd state with $Z_{2}^{E}=+1$, which is most commonly the lightest ordinary
neutralino in this case, is also stable and may account for all or some of the observed cold dark matter density \cite{Athron:2016gor}.

\section{Gauge Coupling Unification}

In this section we consider the RG flow of the gauge couplings within the E$_6$SSM between
$M_Z$ and the GUT scale $M_{X}$. The evolution of these gauge couplings is affected by a kinetic term mixing.
In the Lagrangian of any extension of the SM, that involves an additional $U(1)'$ factor, there can arise a kinetic term
consistent with all symmetries which mixes the gauge fields of the $U(1)'$ and $U(1)_Y$ \cite{Holdom:1985ag}. The E$_6$SSM is not an exception.
In the basis in which the interactions between gauge and matter fields have the canonical form, i.e. for instance a covariant derivative $D_{\mu}$
which acts on the left--handed quark field is given by
\begin{equation}
D_{\mu} = \partial_{\mu} - i g_3 A_{\mu}^a T^a - i g_2 W_{\mu}^b \tau^b - i g_Y Q^{Y}_i B^{Y}_{\mu} - i g_{N} Q^{N}_i B^{N}_{\mu}\,,
\label{17}
\end{equation}
the mixing between the $U(1)$ field strengths can be written as
\begin{equation}
\mathcal{L}_{mix}=-\dfrac{\sin\chi}{2} F^{Y}_{\mu\nu} F^{N}_{\mu\nu}\,.
\label{18}
\end{equation}
Here $A_{\mu}^a$, $W_{\mu}^b$, $B^{Y}_{\mu}$ and $B^{N}_{\mu}$ represent $SU(3)_C$, $SU(2)_W$, $U(1)_Y$ and $U(1)_{N}$ gauge fields; $G_{\mu\nu}^a$, $W_{\mu\nu}^b$, $F_{\mu\nu}^Y$ and $F_{\mu\nu}^{N}$ are field strengths for the corresponding gauge interactions, whereas $g_3$, $g_2$, $g_Y$ and
$g_{N}$ are the $SU(3)_C$, $SU(2)_W$, $U(1)_Y$ and $U(1)_{N}$ gauge couplings in this basis. Since $U(1)_Y$ and $U(1)_{N}$ factors come from the breakdown
of the simple gauge group $E_6$ the parameter $\sin\chi$ is expected to vanish at tree--level. However the non-zero value of this parameter is induced
by loop corrections because
\begin{equation}
Tr\left(Q^YQ^{N}\right)=\sum_{i}\left(Q^Y_i Q^{N}_i\right)\ne 0\,.
\label{19}
\end{equation}
In Eq.~(\ref{19}) trace is restricted to the states which are lighter than $M_X$. The contribution of the complete $E_6$ supermultiplets to this trace
cancels. The non--zero value of the trace (\ref{19}) is induced by $L_4$ and $\overline{L}_4$ supermultiplets which survive to low energies.

For non--zero values of the parameter $\sin\chi$ the mixing in the gauge kinetic part of the Lagrangian (\ref{18}) can be eliminated by means of
a non--unitary transformation of the two $U(1)$ gauge fields \cite{Langacker:1998tc},\cite{Babu:1996vt,Babu:1997st,Rizzo:1998ut,Suematsu:1998wm}:
\begin{equation}
B^Y_{\mu}=B_{1\mu}-B_{2\mu}\tan\chi\,,\qquad B^{N}_{\mu}=B_{2\mu}/\cos\chi\,.
\label{20}
\end{equation}
In the basis $(B_{1\mu}, B_{2\mu})$ the gauge kinetic part of the Lagrangian is diagonal and the covariant derivative (\ref{17}) becomes
\begin{equation}
D_{\mu}=\partial_{\mu}-ig_3A_{\mu}^aT^a-ig_2W_{\mu}^b\tau^b-ig_1
Q^{Y}_iB_{1\mu}-i(g'_1Q^{N}_i+g_{11}Q^Y_i)B_{2\mu}\,,
\label{21}
\end{equation}
where the redefined gauge coupling constants, written in terms of the original ones, are
\begin{equation}
g_1=g_Y\,, \qquad g'_1=g_{N}/\cos\chi\,,\qquad g_{11}=-g_Y\tan\chi\,.
\label{22}
\end{equation}

In the Lagrangian written in terms of the new gauge variables $B_{1\mu}$ and $B_{2\mu}$ the mixing effect is concealed in the
interaction between the $U(1)_{N}$ gauge field and matter fields. The gauge coupling constant $g'_1$ differs from the original one
and there is a new off--diagonal gauge coupling $g_{11}$. In the new basis the covariant derivative (\ref{21}) can be rewritten
in a compact form
\begin{equation}
D_{\mu}=\partial_{\mu} - i g_3 A_{\mu}^a T^a - i g_2 W_{\mu}^b \tau^b - i Q^{T} G B_{\mu}\,,
\label{23}
\end{equation}
where $Q^T=(Q^Y_i,\,Q^{N}_i)$, $B^{T}_{\mu}=(B_{1\mu},\,B_{2\mu})$ and $G$ is a $2\times 2$ matrix of new gauge couplings (\ref{22})
\begin{equation}
G=\left(
\begin{array}{cc}
g_1 & g_{11}\\[0mm]
0   & g'_1
\end{array}
\right)\,.
\label{24}
\end{equation}

Now all physical phenomena can be examined using the Lagrangian with the modified structure of the extra $U(1)_{N}$ interaction (\ref{21})--(\ref{23}).
In this approximation the gauge kinetic mixing changes effectively the $U(1)_{N}$ charges of the fields to
\begin{equation}
\tilde{Q}_i\equiv Q^{N}_i+Q^{Y}_i\delta ,
\label{25}
\end{equation}
where $\delta=g_{11}/g'_1$ while the $U(1)_Y$ charges remain the same. The effective
$U(1)_{N}$ charges $\tilde{Q}_i$ are scale dependent. The particle spectrum in the basis $B^{T}_{\mu}=(B_{1\mu},\,B_{2\mu})$
depends on the effective $U(1)_{N}$ charges $\tilde{Q}_i$.

The running of four diagonal gauge couplings, i.e. $g_3(t)$, $g_2(t)$, $g_1(t)$ and $g'_1(t)$, and one off--diagonal
gauge coupling $g_{11}$ is described by a system of RG equations (RGEs) which can be written in the following form:
\begin{equation}
\dfrac{d G}{d t}=G\times B\,,\qquad\qquad
\dfrac{d g_2}{dt}=\dfrac{\beta_2 g_2^3}{32\pi^2}\,,\qquad\qquad
\dfrac{d g_3}{dt}=\dfrac{\beta_3 g_3^3}{32\pi^2}\,,
\label{26}
\end{equation}
where $t=2\ln\left(q/M_Z\right)$, $q$ is a renormalisation scale, G is a $2\times 2$ matrix (\ref{24}) while $B$ is a $2\times 2$ matrix given by
\begin{equation}
B=\dfrac{1}{32\pi^2}
\left(
\begin{array}{cc}
\beta_1 g_1^2 & 2g_1g'_1\beta_{11}+2g_1g_{11}\beta_1\\[0mm]
0 & g^{'2}_1\beta'_1+2g'_1 g_{11}\beta_{11}+g_{11}^2\beta_1
\end{array}
\right)\,.
\label{27}
\end{equation}
In Eqs.~(\ref{26})--(\ref{27}) $\beta_i$ and $\beta_{11}$ are beta functions.
Here the RG flow of the gauge couplings is explored in the two--loop approximation.
In this approximation $\beta_i$ and $\beta_{11}$ can be presented as a sum of one--loop and two--loop contributions.
In the case of diagonal gauge couplings one gets
\begin{equation}
\beta_i=b_i+\dfrac{\tilde{b}_i}{4\pi}\,,
\label{28}
\end{equation}

It seems to be rather natural to expect that just after the breakdown of the $E_6$ symmetry near the GUT scale $M_X$
there is no mixing in the gauge kinetic part of the Lagrangian between the field strengths associated with the $U(1)_Y$
and $U(1)_{N}$ gauge interactions, while the $SU(3)_C$, $SU(2)_W$, $U(1)_Y$ and $U(1)_{N}$ gauge interactions are
characterised by a unique $E_6$ gauge coupling $g_0$, i.e.
\begin{equation}
g_3(M_X)=g_2(M_X)=g_1(M_X)=g'_1(M_X)=g_0\,,\qquad\qquad g_{11}(M_X)=0\,.
\label{29}
\end{equation}
The previous analysis performed in \cite{King:2007uj} revealed that $g_{11}$ being set to zero at the scale $M_X$ remains
very small at any other scale below $M_X$. Thus it tends to be substantially smaller than the diagonal gauge couplings.
Therefore the two--loop corrections to the off--diagonal beta function $\beta_{11}$ can be neglected.
The one--loop off--diagonal beta function is given by $\beta_{11}=-\dfrac{\sqrt{6}}{5}$.

To simplify our analysis here we further assume that the interactions of matter supermultiplets in the E$_6$SSM are
described by the superpotential (\ref{14}) in which all interactions in $W_{L_4}$ can be ignored,
$\tilde{f}_{\alpha\beta}\simeq f_{\alpha\beta}\to 0$, $\lambda_{\alpha\beta}=\lambda_{\alpha}\delta_{\alpha\beta}$ and
$\kappa_{ij}=\kappa_{i}\delta_{ij}$. The part of the superpotential (\ref{14}) associated with $W_{\rm MSSM}(\mu=0)$
reduces to
\begin{equation}
W_{\rm MSSM}(\mu=0) = h_t Q_3 u^c_3 H_u + h_b Q_3 d^c_3 H_d + h_{\tau} L_3 e^c_3 H_d \,,
\label{30}
\end{equation}
because only third generation fermions have Yukawa couplings to $H_d$ and $H_u$ which can be of the order of unity.
In Eqs.~(\ref{30}) $h_t$, $h_b$ and $h_{\tau}$ are top quark, $b$-quark and $\tau$--lepton Yukawa couplings respectively.

In the one--loop approximation the beta functions of the diagonal gauge couplings are given by
\begin{equation}
b_1=\dfrac{3}{5} + 3 N_g\,,\qquad b'_1 = \dfrac{2}{5}+3N_g\,,\qquad b_2=-5+3N_g\,,\qquad b_3=-9+3N_g\,,
\label{31}
\end{equation}
where parameter $N_g$ is the number of generations in the E$_6$SSM forming complete $E_6$ fundamental representations
at low energies ($E<<M_X$). As one can see $N_g=3$ is the critical value for the one--loop beta function of the strong interactions.
Since $N_g=3$ in the E$_6$SSM $b_3$ is equal to zero and in the one--loop approximation the $SU(3)_C$ gauge coupling remains constant
everywhere from the EW scale to $M_X$. Thus any reliable analysis of gauge coupling unification requires the inclusion of
two--loop corrections to the beta functions of the diagonal gauge couplings in the E$_6$SSM. Using the results
of the computation of two--loop beta functions in a general softly broken $N=1$ SUSY model \cite{Martin:1993zk} one obtains
\begin{equation}
\begin{array}{rcl}
\tilde{b}_1&=& 8N_g \alpha_3+\left(\dfrac{9}{5}+3N_g\right)\alpha_2+\left(\dfrac{9}{25}+3 N_g\right) \alpha_1+
\left(\dfrac{6}{25}+N_g\right) \alpha'_1\\[0mm]
&&-\dfrac{26}{5} y_t-\dfrac{14}{5} y_b-\dfrac{18}{5} y_{\tau}-\dfrac{6}{5}\Sigma_{\lambda}-
\dfrac{4}{5}\Sigma_{\kappa}\,,\\[0mm]
\tilde{b}'_1&=& 8N_g \alpha_3+\left(\dfrac{6}{5}+3N_g\right)\alpha_2+
\left(\dfrac{6}{25}+ N_g\right) \alpha_1+\left(\dfrac{4}{25}+3N_g\right) \alpha'_1\\[3mm]
&&-\dfrac{9}{5} y_t-\dfrac{21}{5}y_b-\dfrac{7}{5}y_{\tau}-\dfrac{19}{5}\Sigma_{\lambda}-
\dfrac{57}{10}\Sigma_{\kappa}\,,\\[0mm]
\tilde{b}_2&=&8N_g \alpha_3+\biggl(-17+21 N_g\biggr)\alpha_2+ \left(\dfrac{3}{5}+N_g\right) \alpha_1+
\left(\dfrac{2}{5}+N_g\right) \alpha'_1 \\[0mm]
&&- 6 y_t-6 y_b-2 y_{\tau}-2\Sigma_{\lambda}\,,\\[0mm]
\tilde{b}_3&=&\alpha_3\biggl(-54+34 N_g\biggr)+3 N_g \alpha_2+ N_g \alpha_1+N_g \alpha'_1-4 y_t-4 y_b- 2\Sigma_{\kappa}\,,\\[0mm]
\Sigma_{\lambda}&=&y_{\lambda_1}+y_{\lambda_2}+y_{\lambda}\,,\qquad\qquad\qquad\Sigma_{\kappa}=y_{\kappa_1}+y_{\kappa_2}+y_{\kappa_3}\,,
\end{array}
\label{32}
\end{equation}
where $\alpha_i=\dfrac{g_i^2}{4\pi}$, $\alpha'_1=\dfrac{g_1^{'2}}{4\pi}$, $y_t=\dfrac{h_t^2}{4\pi}$,\, $y_b=\dfrac{h_b^2}{4\pi}$,\, $y_{\tau}=\dfrac{h_{\tau}^2}{4\pi}$,\, $y_{\lambda}=\dfrac{\lambda^2}{4\pi}$,\, $y_{\lambda_{\alpha}}=\dfrac{\lambda_{\alpha}^2}{4\pi}$ and $y_{\kappa_i}=\dfrac{\kappa_i^2}{4\pi}$.

For the analysis of the RG flow of the SM gauge couplings it is convenient to use an approximate solution of the two--loop RGEs
(see \cite{Chankowski:1995dm}). At high energies this solution can be written as
\begin{equation}
\dfrac{1}{\alpha_i(t)}=\dfrac{1}{\alpha_i(M_Z)}-\dfrac{b_i}{2\pi} t-\frac{C_i}{12\pi}-\Theta_i(t)
+\dfrac{b_i-b_i^{SM}}{2\pi}\ln\frac{T_i}{M_Z}\,,
\label{33}
\end{equation}
where $b_i^{SM}$ are the coefficients of the one--loop beta functions in the SM, the third term in the
right--hand side of Eq.~(\ref{3}) is the $\overline{MS}\to\overline{DR}$ conversion factor with $C_1=0$, $C_2=2$, $C_3=3$ \cite{Antoniadis:1982vr,Antoniadis:1982qw}, while
\begin{equation}
\Theta_i(t)=\dfrac{1}{8\pi^2}\int_0^t \tilde{b}_i d\tau\,,\qquad\qquad
T_i=\prod_{k=1}^N\biggl(m_k\biggr)^{\dfrac{\Delta b^k_i}{b_i-b_i^{SM}}}\,
\label{34}
\end{equation}
In Eq.~(\ref{34}) $m_k$ and $\Delta b_i^k $ are masses and one--loop contributions to $b_i$ due to new particles appearing in the E$_6$SSM.
Since the two--loop corrections to the running of the gauge couplings $\Theta_i(t)$ are considerably smaller than the leading terms,
the solutions of the one--loop RGEs for the gauge and Yukawa couplings are normally used for the calculation of $\Theta_i(t)$.
The threshold corrections associated with the last terms in Eq.~(\ref{33}) are of the same order as or even less than $\Theta_i(t)$.
Therefore in Eqs.~(\ref{33})--(\ref{34}) only one--loop threshold effects are taken into account.

Relying on the approximate solution of the two--loop RGEs one can find the relationships between the values of the gauge couplings at
low energies and GUT scale. Then by using the expressions describing the RG flow of $\alpha_1(t)$ and $\alpha_2(t)$ one can estimate
the scale $M_X$ where $\alpha_1(M_X)=\alpha_2(M_X)=\alpha_0$ and the value of the overall gauge coupling $\alpha_0$ at this scale.
Substituting $M_X$ and $\alpha_0$ into the solution of the RGE for the strong gauge coupling the value of $\alpha_3(M_Z)$, for
which exact gauge coupling unification takes place, may be obtained (see \cite{Carena:1993ag}):
\begin{equation}
\begin{array}{c}
\dfrac{1}{\alpha_3(M_Z)}=\dfrac{1}{b_1-b_2}\biggl[\dfrac{b_1-b_3}{\alpha_2(M_Z)}-
\dfrac{b_2-b_3}{\alpha_1(M_Z)}\biggr]-\dfrac{1}{28\pi}+\Theta_s+\dfrac{19}{28\pi}\ln\dfrac{T_{S}}{M_Z}\,,\\[0mm]
\Theta_s=\biggl(\dfrac{b_2-b_3}{b_1-b_2}\Theta_1-\dfrac{b_1-b_3}{b_1-b_2}\Theta_2+\Theta_3\biggr)\,,\qquad \Theta_i=\Theta_i(M_X)\,.
\end{array}
\label{35}
\end{equation}
The combined threshold scale $T_{S}$ can be expressed in terms of the effective threshold scales $T_1$, $T_2$ and $T_3$
\begin{equation}
T_{S}=\dfrac{T_2^{172/19}}{T_1^{55/19} T_3^{98/19}}\,.
\label{36}
\end{equation}
In Eq.~(\ref{36}) $T_1$, $T_2$ and $T_3$ are given by
\begin{equation}
\begin{array}{rcl}
T_1&=&\mu^{4/55} m_{A}^{1/55} \mu_{L}^{4/55} m_{L}^{2/55}
\Biggl(\prod_{i=1,2,3} m_{\tilde{Q}_i}^{1/165} m_{\tilde{d}_i}^{2/165} m_{\tilde{u}_i}^{8/165}
m_{\tilde{L}_i}^{1/55} m_{\tilde{e}_i}^{2/55} m_{\tilde{D}_i}^{4/165}\mu_{D_i}^{8/165}\Biggr)\times\\[0mm]
&&\times\Biggl(\prod_{\alpha=1,2}m_{H_{\alpha}}^{2/55}\mu_{\tilde{H}_{\alpha}}^{4/55}\Biggr)\,,\\[0mm]
T_2&=& M_{\tilde{W}}^{8/43} \mu^{4/43} m_A^{1/43} \mu_{L}^{4/43} m_{L}^{2/43}
\Biggl(\prod_{i=1,2,3} m_{\tilde{Q}_i}^{3/43} m_{\tilde{L}_i}^{1/43}\Biggr)
\Biggl(\prod_{\alpha=1,2} m_{H_{\alpha}}^{2/43}\mu_{\tilde{H}_{\alpha}}^{4/43}\Biggr)\,,\\[0mm]
T_3&=&M_{\tilde{g}}^{2/7} \Biggl(\prod_{i=1,2,3} m_{\tilde{Q}_i}^{1/21} m_{\tilde{u}_i}^{1/42} m_{\tilde{d}_i}^{1/42} m_{\tilde{D}_i}^{1/21}\mu_{D_i}^{2/21}\Biggr)\,,
\end{array}
\label{37}
\end{equation}
where $M_{\tilde{g}}$ and $M_{\tilde{W}}$ are the masses of gluinos and winos; $\mu$ and $m_A$ are effective $\mu$--term and
the masses of heavy Higgs states respectively; $m_{\tilde{u}_i}$, $m_{\tilde{d}_i}$ and $m_{\tilde{Q}_i}$ are the masses of the
right--handed and left--handed squarks; $m_{\tilde{L}_i}$ and $m_{\tilde{e}_i}$ are the masses of the left--handed and right--handed
sleptons; $m_{H_{\alpha}}$ and $\mu_{\tilde{H}_{\alpha}}$ are the masses of the scalar and fermion components of $H^u_{\alpha}$ and $H^d_{\alpha}$;
$\mu_{D_i}$ and $m_{\tilde{D}_i}$ are the masses of exotic quarks and their superpartners; $m_{L}$ and $\mu_{L}$ are the masses of the scalar
and fermion components of $L_4$ and $\overline{L}_4$.

It is worth noting here that in the limit when the two--loop and threshold corrections are neglected, i.e.
$\Theta_s=0$ and $T_{S}=M_Z$, Eq.~(\ref{33}) leads to the same prediction for $\alpha_3(M_Z)$ in the MSSM and E$_6$SSM.
Indeed, since extra matter in the E$_6$SSM form complete $SU(5)$ representations these multiplets contribute equally
to the one--loop beta functions of the $SU(3)_C$, $SU(2)_W$ and $U(1)_Y$ interactions. Due to this the differences
of the coefficients of the one--loop beta functions $b_i-b_j$ and the form of Eq.~(\ref{33}) remain the same
in the MSSM and E$_6$SSM. However the inclusion of the two--loop and threshold corrections may spoil the unification
of the SM gauge couplings entirely within the E$_6$SSM.

\begin{figure}
\centering
\includegraphics[width=7.7cm]{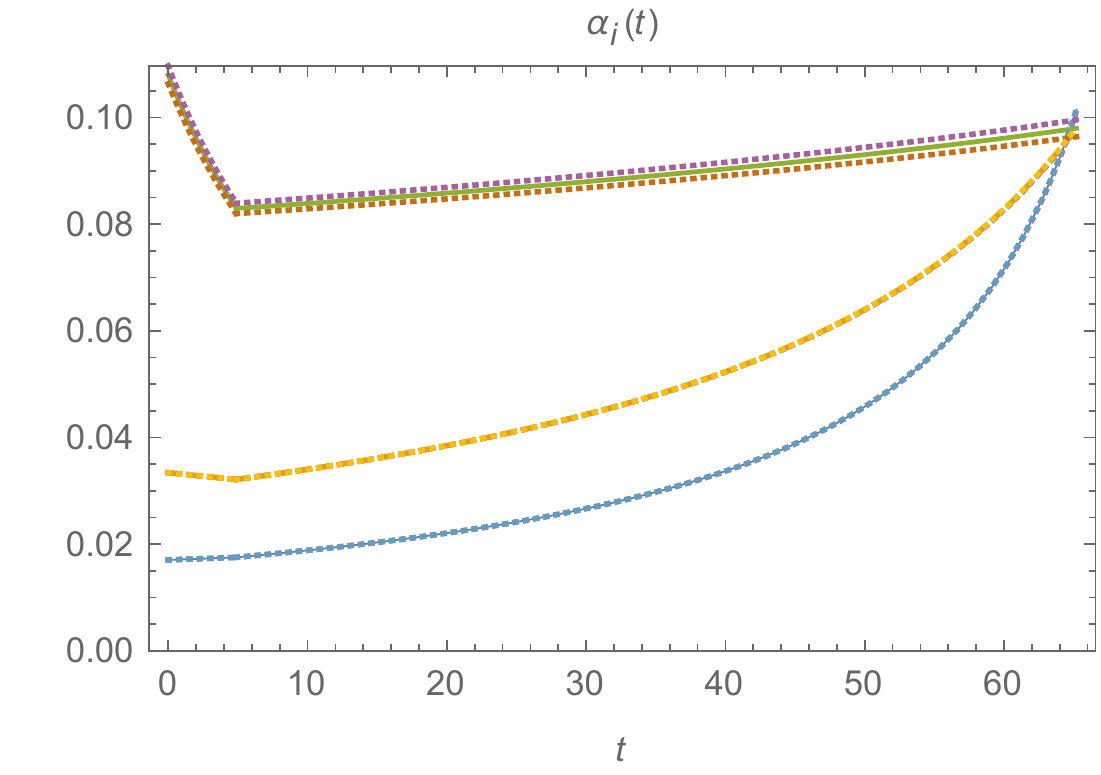}
\includegraphics[width=7.7cm]{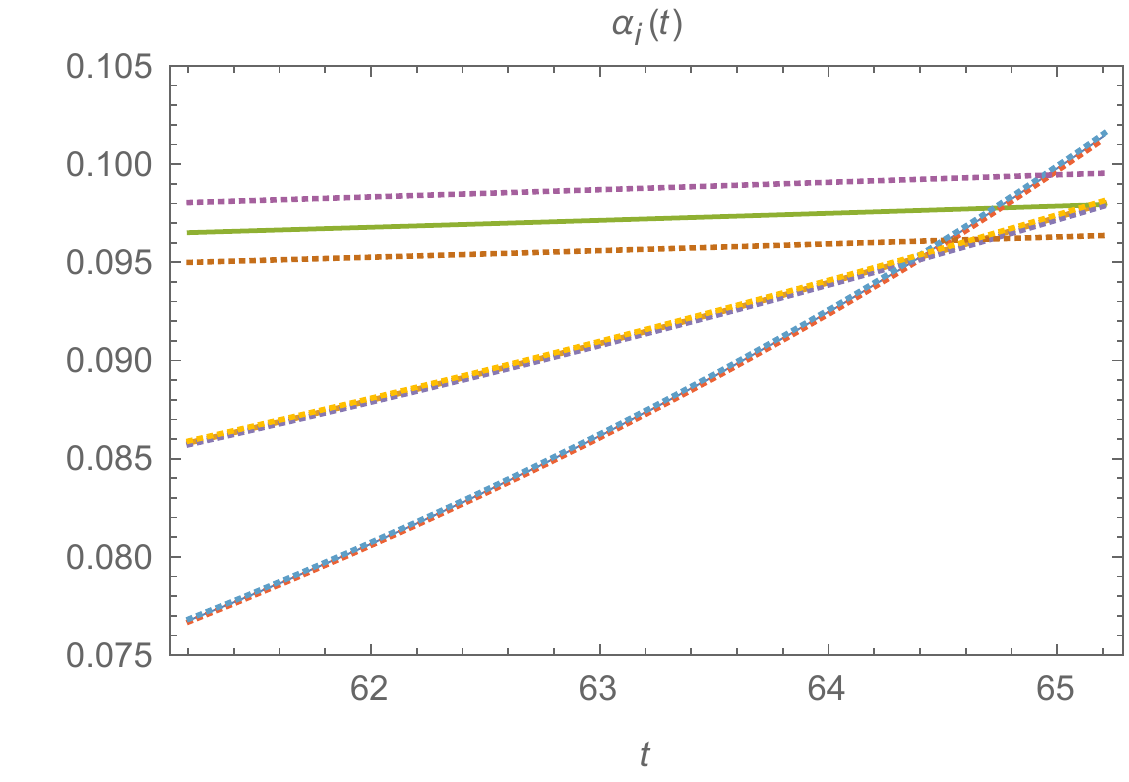}
\caption{{\it Left:} Two--loop RG flow of $SU(3)_C$, $SU(2)_W$ and $U(1)_Y$ couplings
in the E$_6$SSM as a function of $t=2\ln\left(q/M_Z\right)$ for $T_S=2\,\mbox{TeV}$
in the case when $q$ varies from $M_Z$ to $M_X$.
{\it Right:} Evolution of these couplings in the vicinity of $M_X$.
Thick, dashed and solid lines
correspond to the running of $SU(3)_C$, $SU(2)_W$ and $U(1)_Y$ couplings
respectively. We used $\tan\beta=10$, $\alpha_s(M_Z)=0.118$,
$\alpha(M_Z)=1/127.9$, $\sin^2\theta_W=0.231$ and $\kappa_1(T_S)=\kappa_2(T_S)
=\kappa_3(T_S)=\lambda_1(T_S)=\lambda_2(T_S)=\lambda(T_S)=g^{'}_1(T_S)$.
The dotted lines represent the uncertainty in $\alpha_i(t)$ caused by
the variation of the strong gauge coupling from 0.116 to 0.120 at the EW scale.}
\label{essmfig1}
\end{figure}

In general $T_1$, $T_2$ and $T_3$ in Eq.~(\ref{37}) can be quite different. Nevertheless from Eq.~(\ref{35}) it follows that the unification of
the SM gauge couplings is determined by a single combined threshold scale $T_{S}$. Therefore without loss of generality one can set
three effective threshold scales be equal to each other. Then from Eq.~(\ref{36}) it follows that $T_1=T_2=T_3=T_S$. The results
of our numerical analysis of the gauge coupling unification within the E$_6$SSM are presented in Figure~\ref{essmfig1} where
the two--loop RG flow of gauge couplings is shown. We use the two--loop SM beta functions to evaluate the running of gauge
couplings between $M_Z$ and $T_1=T_2=T_3=T_S$. Then we apply the two--loop RGEs of the E$_6$SSM to calculate the evolution
of $\alpha_i(t)$ from $T_S$ to $M_X$ which is around $2-3\cdot 10^{16}\,\mbox{GeV}$ in the case of the E$_6$SSM. The low energy
values of $g'_1$ and $g_{11}$ are chosen so that the conditions (\ref{29}) are fulfilled. For the computation of the RG flow
of Yukawa couplings a set of one--loop RGEs is used. The corresponding one--loop RGEs are specified in \cite{King:2005jy}.

In Figure~\ref{essmfig1} we fix $T_1=T_2=T_3=T_S=2\,\mbox{TeV}$ and $\tan\beta=10$.
Although to simplify our analysis we also set $\kappa_i(T_S)=\lambda_{\alpha}(T_S)=\lambda(T_S)=g^{'}_1(T_S)$ the RG flow
of $\alpha_i(t)$ depends rather weakly on the values of the Yukawa and extra $U(1)_N$ gauge couplings.
Dotted lines in Figure~\ref{essmfig1} show the changes of the evolution of gauge couplings induced by the variations of
$\alpha_3(M_Z)$ from $0.116$ to $0.120$. The corresponding interval of variations of $\alpha_3(t)$ is
always considerably wider than the ones for $\alpha_1(t)$ and $\alpha_2(t)$. The dependence of $\alpha_1(t)$ and $\alpha_2(t)$
on the value of the strong gauge coupling at the EW scale is expected to be relatively weak because $\alpha_3(t)$ appears
only in the two--loop contributions to $\beta_1$ and $\beta_2$. It is worthwhile to point out that at high energies
the uncertainty in $\alpha_3(t)$ caused by the variations of $\alpha_3(M_Z)$ is much bigger in the E$_6$SSM than in the MSSM.
This happens because in the E$_6$SSM the strong gauge coupling grows with increasing renormalisation scale $q$ whereas in the MSSM
it decreases at high energies. Thus the uncertainty in $\alpha_3(M_X)$ in the E$_6$SSM is approximately equal
to the low energy uncertainty in $\alpha_3(M_Z)$ while in the MSSM the interval of variations of $\alpha_3(M_X)$ shrinks drastically.
As a consequence it is much easier to achieve the unification of gauge couplings within the E$_6$SSM as compared with the MSSM
where in the two--loop approximation the exact gauge coupling unification requires $\alpha_3(M_Z)>0.123$, well above the experimentally
measured central value \cite{Chankowski:1995dm}, \cite{Carena:1993ag},
\cite{Bagger:1995bw,Langacker:1995fk,Langacker:1992rq,gc-1,gc-2,gc-3,deBoer:2003xm,deBoer:2005bd}.

The results of the numerical analysis presented in Figure~\ref{essmfig1} demonstrate that for $T_S=2\,\mbox{TeV}$
almost exact unification of the SM gauge couplings can be achieved in the E$_6$SSM if $\alpha_3(M_Z)\approx 0.116$.
With increasing (decreasing) the effective threshold scale $T_S$ the value of $\alpha_3(M_Z)$, at which the exact
gauge coupling unification takes place, becomes lower (greater). In the E$_6$SSM $T_S$ can be considerably lower
than $1\,\mbox{TeV}$ even when the SUSY breaking scale is much larger than a few TeV. To demonstrate this let us
assume that all scalars except the SM--like Higgs boson are almost degenerate around $m_A\approx M_S$ which is much larger
than the masses of all fermions. Then combining Eqs.~(\ref{36})--(\ref{37}) one finds
\begin{equation}
T_S=\dfrac{M_{\tilde{W}}^{32/19} M_S^{3/19}}{M_{\tilde{g}}^{28/19}}
\left(\dfrac{\mu \mu_{L} \mu_{\tilde{H}_{1}} \mu_{\tilde{H}_{2}}}{\mu_{D_1}\mu_{D_2}\mu_{D_3}}\right)^{12/19}\,.
\label{38}
\end{equation}
If $M_S\approx M_{\tilde{g}} \approx 10\,\mbox{TeV}$ and
$\mu\approx \mu_{L}\approx \mu_{\tilde{H}_{1}}\approx \mu_{\tilde{H}_{2}}\approx 1\,\mbox{TeV}$ while $M_{\tilde{W}}$ and
the masses of the exotic quarks $\mu_{D_i}$ are of the order of a few TeV, the effective threshold scale tends to be
much smaller than $1\,\mbox{TeV}$. For $T_S=400\,\mbox{GeV}$ almost exact unification of the SM gauge couplings in the E$_6$SSM
can be obtained if $\alpha_3(M_Z)\approx 0.118$ \cite{Nevzorov:2012hs}. Thus in this SUSY model the gauge coupling unification
can be attained for the values of $\alpha_3(M_Z)$ which are in agreement with current data.

As was mentioned before the inclusion of the two--loop corrections to the diagonal beta functions could spoil
the unification of the SM gauge couplings entirely within the E$_6$SSM. These corrections affect the running
of gauge couplings much more strongly than in the case of the MSSM because at any intermediate scale the values
of the gauge couplings in the E$_6$SSM are substantially larger as compared to the ones in the MSSM.
The analysis of the RG flow of the SM gauge couplings performed in \cite{King:2007uj} revealed that
$\Theta_i(M_X)$ are a few times larger in the E$_6$SSM than in the MSSM. On the other hand due to the remarkable
cancellation of different two--loop corrections the absolute value of $\Theta_s$ is more than three times
smaller in the E$_6$SSM as compared with the MSSM. Such cancellation is caused by the structure of the two--loop corrections
to the diagonal beta functions in the model under consideration. As a result, the prediction for $\alpha_3(M_Z)$
obtained using Eq.~(\ref{35}) is considerably lower in the E$_6$SSM than in the MSSM.

\section{Gauge symmetry breaking and Higgs sector}

In the simplest case the sector responsible for the breakdown of the $SU(2)_W\times U(1)_Y\times U(1)_{N}$ gauge
symmetry in the E$_6$SSM involves two Higgs doublets $H_u$ and $H_d$ as well as the SM singlet field $S$.
The interactions between these fields are determined by the structure of the gauge group and by the superpotential (\ref{14}).
Including soft SUSY breaking terms and radiative corrections, the resulting Higgs effective potential
is the sum of four pieces:
\begin{equation}
V=V_F+V_D+V_{\rm soft}+\Delta V\, ,
\label{39}
\end{equation}
\begin{equation}
V_F=\lambda^2 |S|^2 (|H_d|^2+|H_u|^2) + \lambda^2 |(H_d H_u)|^2\,,
\label{40}
\end{equation}
\begin{equation}
\begin{array}{rcl}
V_D&=&\sum_{a=1}^3 \dfrac{g_2^2}{8}\left(H_d^\dagger \sigma_a H_d+H_u^\dagger \sigma_a H_u\right)^2
+\dfrac{{g'}^2}{8}\left(|H_d|^2-|H_u|^2\right)^2+\\
&&+\dfrac{g^{'2}_1}{2}\left(\tilde{Q}_{H_d} |H_d|^2 + \tilde{Q}_{H_u} |H_u|^2 + \tilde{Q}_S|S|^2\right)^2\,,
\end{array}
\label{41}
\end{equation}
\begin{equation}
V_{\rm soft}=m_{S}^2|S|^2+m_{H_d}^2|H_d|^2+m_{H_u}^2|H_u|^2+\biggl[\lambda A_{\lambda}S(H_u H_d)+{\rm h.c.}\biggr]\,,
\label{42}
\end{equation}
where $\sigma_a$ ($a=1,2,3$) denote the three Pauli matrices, $g'=\sqrt{3/5} g_1$, $H_d^T=(H_d^0,\,H_d^{-})$,
$H_u^T=(H_u^{+},\,H_u^{0})$, $(H_d H_u)=H_u^{+}H_d^{-}-H_u^{0}H_d^{0}$, while $\tilde{Q}_{H_d}$, $\tilde{Q}_{H_u}$,
and $\tilde{Q}_S$ are the effective  $U(1)_{N}$ charges of $H_d$, $H_u$ and $S$, respectively.
At tree-level the Higgs potential in Eq.~(\ref{39}) is described by the sum of the first three terms.
$V_F$ and $V_D$ contain the $F$-and $D$-term contributions that do not violate SUSY.
The structure of the $F$--terms $V_F$ is exactly the same as in the NMSSM without the self--interaction of the SM singlet superfield $S$.
However the $D$--terms in $V_D$ contain a new ingredient: the terms in the expression for $V_D$ proportional to ${g'_1}^2$
represent $D$--term contributions due to the extra $U(1)_{N}$ which are not present in the MSSM or NMSSM.
The low--energy value of the extra $U(1)_{N}$ coupling $g'_1$ and the effective $U(1)_{N}$ charges of $H_d$, $H_u$ and $S$
can be computed assuming gauge coupling unification whereas $g_2$ and $g'$ are well known.
The soft SUSY breaking terms are collected in $V_{soft}$ and include the soft masses
$m_{H_d}^2,\,m_{H_u}^2,\,m_{S}^2$ as well as trilinear coupling $A_{\lambda}$. This part of the scalar potential (\ref{39})
coincides with the corresponding one in the NMSSM when the NMSSM parameters $\kappa$ and $A_{\kappa}$ vanish.
Because the only complex phase (of $\lambda A_{\lambda}$) that appears in the tree--level scalar potential (\ref{39})
can easily be absorbed by a suitable redefinition of the Higgs fields, CP--invariance is preserved in the Higgs sector of
the E$_6$SSM at tree--level.

The term $\Delta V$ represents the contribution of loop corrections to the Higgs effective potential.
In the one--loop approximation the contributions of different states to $\Delta V$ are determined
by their masses, i.e.
\begin{equation}
\Delta V = \dfrac{1}{64\pi^2} Str\,|M|^4\biggl[\log\dfrac{|M|^2}{Q^2} - \dfrac{3}{2} \biggr],
\label{43}
\end{equation}
where $M$ is the mass matrix for the bosons and fermions in the SUSY model under consideration.
Here the supertrace operator counts positively (negatively) the number of degrees of freedom for the different
bosonic (fermionic) fields, while $Q$ is the renormalisation scale. The inclusion of loop corrections draws into
the analysis many other soft SUSY breaking parameters which determine masses of different superparticles.
Some of these parameters may be complex giving rise to  potential sources of CP--violation.

At the physical minimum of the scalar potential (\ref{39}) the Higgs fields develop VEVs
\begin{equation}
<H_d>=\dfrac{1}{\sqrt{2}}\left(
\begin{array}{c}
v_1\\ 0
\end{array}
\right) , \qquad
<H_u>=\dfrac{1}{\sqrt{2}}\left(
\begin{array}{c}
0\\ v_2
\end{array}
\right) ,\qquad
<S>=\dfrac{s}{\sqrt{2}}.
\label{44}
\end{equation}
The equations for the extrema of the full Higgs boson effective potential in the
directions (\ref{44}) in field space read:
\begin{equation}
\dfrac{\partial V}{\partial s} = m_{S}^2 s - \dfrac{\lambda A_{\lambda}}{\sqrt{2}}v_1v_2 + \dfrac{\lambda^2}{2}(v_1^2+v_2^2)s
+ \dfrac{g^{'2}_1}{2} D' \tilde{Q}_S s +\dfrac{\partial\Delta V}{\partial s}=0\,,
\label{45}
\end{equation}
\begin{equation}
\dfrac{\partial V}{\partial v_1} = m_{H_d}^2 v_1 - \dfrac{\lambda A_{\lambda}}{\sqrt{2}} s v_2 + \dfrac{\lambda^2}{2}(v_2^2+s^2)v_1
+ \dfrac{\bar{g}^2}{8}\biggl(v_1^2-v_2^2)\biggr)v_1 + \dfrac{g^{'2}_1}{2} D' \tilde{Q}_{H_d} v_1 + \dfrac{\partial\Delta V}{\partial v_1}=0\,,
\label{46}
\end{equation}
\begin{equation}
\dfrac{\partial V}{\partial v_2} = m_{H_u}^2 v_2 - \dfrac{\lambda A_{\lambda}}{\sqrt{2}} s v_1 + \dfrac{\lambda^2}{2}(v_1^2+s^2)v_2 + \dfrac{\bar{g}^2}{8}\biggl(v_2^2-v_1^2\biggr)v_2 + \dfrac{g^{'2}_1}{2} D' \tilde{Q}_{H_u} v_2 + \dfrac{\partial\Delta V}{\partial v_2}=0\,,
\label{47}
\end{equation}
where $D'=\tilde{Q}_{H_d} v_1^2 + \tilde{Q}_{H_u}v_2^2 + \tilde{Q}_S s^2$ and $\bar{g}=\sqrt{g_2^2+g'^2}$.
Instead of specifying $v_1$ and $v_2$ it is more convenient to use $v=\sqrt{v_1^2+v_2^2} \approx 246\,\mbox{GeV}$ and $\tan\beta=v_2/v_1$.

The Higgs sector of the E$_6$SSM includes ten degrees of freedom. Four of them are massless Goldstone modes
which are swallowed by the $W^{\pm}$, $Z$ and $Z'$ gauge bosons. The charged $W^{\pm}$ bosons gain masses
via the interaction with the neutral components of the Higgs doublets $H_u$ and $H_d$ just in the same way as in the
MSSM, resulting in $M_W=\dfrac{g_2}{2}v$. Meanwhile the mechanism of the neutral gauge boson mass generation differs substantially.
Letting the $Z'$ and $Z$ states be the gauge bosons associated with $U(1)_{N}$ and with the SM-like $Z$ boson
the $Z-Z'$ mass squared matrix is given by
\begin{equation}
M^2_{ZZ'}=\left(
\begin{array}{cc}
\dfrac{\bar{g}^2}{4} v^2  &  \dfrac{\bar{g}g'_1}{2}v^2\biggl(\tilde{Q}_{H_d}\cos^2\beta-\tilde{Q}_{H_u}\sin^2\beta\biggr)\\[0mm]
\dfrac{\bar{g}g'_1}{2}v^2\biggl(\tilde{Q}_{H_d}\cos^2\beta-\tilde{Q}_{H_u}\sin^2\beta\biggr)  & g^{\prime \,2}_1 D'
\end{array}
\right)\,.
\label{48}
\end{equation}
The SM singlet fields $S$ must acquire large VEV, $s \gg 1\,\mbox{TeV}$, to ensure
that the extra $U(1)_{N}$ gauge boson is sufficiently heavy. In this case the mass of the lightest neutral gauge boson $Z_1$ is very close
to $M_Z=\bar{g}v/2$, while the mass of $Z_2$ is set by the VEV of the SM singlet field, i.e. $M_{Z'}\approx  g'_1\tilde{Q}_S\, s$.

For the analysis of the spectrum of the Higgs bosons in the E$_6$SSM we use Eq.~(\ref{45})--(\ref{47})
for the extrema to express the soft masses $m_{H_d}^2,\,m_{H_u}^2,\,m_{S}^2$ in terms of $s,\, v$, $\tan\beta$ and other parameters.
Because of the conversation of the electric charge, the charged components of the Higgs doublets are not mixed with the neutral Higgs fields.
They form a separate sector, whose spectrum is described by a $2\times 2$ mass matrix.
The determinant of this matrix vanishes leading to the appearance of two Goldstone states
\begin{equation}
G^-=H_d^{-}\cos\beta-H_u^{+*}\sin\beta
\label{49}
\end{equation}
and its charge conjugate which are absorbed into the longitudinal degrees of freedom of the $W^{\pm}$ gauge boson.
Their orthogonal linear combination
\begin{equation}
H^{+}=H_d^{-*}\sin\beta+H_u^{+}\cos\beta
\label{50}
\end{equation}
gains mass
\begin{equation}
m^2_{H^{\pm}}=\dfrac{\sqrt{2}\lambda A_{\lambda}}{\sin 2\beta}s
-\dfrac{\lambda^2}{2}v^2+\dfrac{g^2}{2}v^2+\Delta_{\pm}\,.
\label{51}
\end{equation}
where $\Delta_{\pm}$ denotes the loop corrections to $m^2_{H^{\pm}}$.

If CP-invariance is preserved then the imaginary parts of the neutral components of the Higgs doublets and the SM singlet field $S$
do not mix with the real parts of these fields. In this case the imaginary parts of the neutral components of the Higgs doublets
as well as imaginary part of the SM singlet field $S$ form CP--odd Higgs sector. They compose two neutral Goldstone states
\begin{equation}
\begin{array}{l}
G=\sqrt{2}(\mbox{Im}\,H_d^0 \cos\beta - \mbox{Im}\, H_u^0 \sin\beta)\,,\\[0mm]
G'=\sqrt{2}\mbox{Im}\,S \cos\gamma - \sqrt{2}(\mbox{Im}\,H_u^0\cos\beta + \mbox{Im}\,H_d^0\sin\beta)\sin\gamma\,,
\end{array}
\label{52}
\end{equation}
which are swallowed by the $Z$ and $Z'$ bosons, and one physical state
\begin{equation}
A=\sqrt{2}\mbox{Im}\,S \sin\gamma + \sqrt{2}(\mbox{Im}\,H_u^0\cos\beta + \mbox{Im}\,H_d^0\sin\beta)\cos\gamma\,,
\label{53}
\end{equation}
where $\tan\gamma=\dfrac{v}{2s}\sin2\beta$. Two massless pseudoscalars $G_0$ and $G'$ decouple from the rest of
the spectrum whereas the physical CP--odd Higgs state $A$ acquires mass
\begin{equation}
m^2_{A}=\dfrac{\sqrt{2}\lambda A_{\lambda}}{\sin 2\gamma}v+\Delta_A\,,
\label{54}
\end{equation}
In Eq.~(\ref{54}) $\Delta_A$ denote loop corrections. Since in the E$_6$SSM $s$ must be much
larger than $v$, the value of $\gamma$ is always small and the physical pseudoscalar is predominantly the superposition
of the imaginary parts of the neutral components of the Higgs doublets. In the limit $s\gg v$ the masses of the charged and
CP--odd Higgs states are approximately equal to each other.

The CP--even Higgs sector includes $Re\,H_d^0$, $Re\,H_u^0$ and $Re\,S$.
In the field space basis $(h,\,H,\,N)$, where
\begin{equation}
\begin{array}{l}
Re\,H_d^0=(h \cos\beta- H \sin\beta+v_1)/\sqrt{2}\,, \\
Re\,H_u^0=(h \sin\beta+ H \cos\beta+v_2)/\sqrt{2}\,, \\
Re\,S=(s+N)/\sqrt{2}\,,
\end{array}
\label{55}
\end{equation}
the mass matrix of the CP-even Higgs sector takes the form \cite{Kovalenko:1998dc,Nevzorov:2000uv,Nevzorov:2001um}:
\begin{equation}
M^2=\left(
\begin{array}{ccc}
M_{11}^2 & M_{12}^2 & M_{13}^2\\
M_{21}^2 & M_{22}^2 & M_{23}^2\\
M_{31}^2 & M_{32}^2 & M_{33}^2
\end{array}
\right)=
\left(
\begin{array}{ccc}
\dfrac{\partial^2 V}{\partial v^2}& \dfrac{1}{v}\dfrac{\partial^2 V}{\partial v \partial\beta}&
\dfrac{\partial^2 V}{\partial v \partial s}\\[0cm]
\dfrac{1}{v}\dfrac{\partial^2 V}{\partial v \partial\beta}& \dfrac{1}{v^2}\dfrac{\partial^2 V}{\partial^2\beta}&
\dfrac{1}{v}\dfrac{\partial^2 V}{\partial s \partial\beta}\\[0cm]
\dfrac{\partial^2 V}{\partial v \partial s}& \dfrac{1}{v}\dfrac{\partial^2 V}{\partial s \partial\beta}&
\dfrac{\partial^2 V}{\partial^2 s}
\end{array}
\right)~.
\label{56}
\end{equation}

Taking second derivatives of the Higgs effective potential (\ref{39})--(\ref{42}) and substituting $m_{H_d}^2$, $m_{H_u}^2$, $m_{S}^2$
from the minimisation conditions (\ref{45})--(\ref{47}) one finds:
\begin{equation}
M_{11}^2=\dfrac{\lambda^2}{2}v^2\sin^22\beta+\dfrac{\bar{g}^2}{4}v^2\cos^22\beta+g^{'2}_1 v^2(\tilde{Q}_{H_d}\cos^2\beta+
\tilde{Q}_{H_u}\sin^2\beta)^2+\Delta_{11}\,,
\label{57}
\end{equation}
\begin{equation}
\begin{array}{rcl}
M_{12}^2=M_{21}^2&=&\left(\dfrac{\lambda^2}{4}-\dfrac{\bar{g}^2}{8}\right)v^2 \sin 4\beta+\dfrac{g^{'2}_1}{2}v^2(\tilde{Q}_{H_u}-\tilde{Q}_{H_d})\times\\
&&\times(\tilde{Q}_{H_d}\cos^2\beta+\tilde{Q}_{H_u}\sin^2\beta)\sin 2\beta+\Delta_{12}\, ,\\
\end{array}
\label{58}
\end{equation}
\begin{equation}
M_{22}^2=\dfrac{\sqrt{2}\lambda A_{\lambda}}{\sin 2\beta}s+\left(\dfrac{\bar{g}^2}{4}-\dfrac{\lambda^2}{2}\right)v^2
\sin^2 2\beta+\dfrac{g^{'2}_1}{4}(\tilde{Q}_{H_u}-\tilde{Q}_{H_d})^2 v^2 \sin^22\beta+\Delta_{22}\,,
\label{59}
\end{equation}
\begin{equation}
M_{23}^2=M_{32}^2=-\dfrac{\lambda A_{\lambda}}{\sqrt{2}}v\cos 2\beta+\dfrac{g^{'2}_1}{2}(\tilde{Q}_{H_u}-\tilde{Q}_{H_d})\tilde{Q}_S v s
\sin 2\beta+\Delta_{23}\,,
\label{60}
\end{equation}
\begin{equation}
M_{13}^2=M_{31}^2=-\dfrac{\lambda A_{\lambda}}{\sqrt{2}}v\sin 2\beta+\lambda^2 v s+g^{'2}_1(\tilde{Q}_{H_d}\cos^2\beta+
\tilde{Q}_{H_u}\sin^2\beta)\tilde{Q}_S v s+\Delta_{13}\,,\\
\label{61}
\end{equation}
\begin{equation}
M_{33}^2=\dfrac{\lambda A_{\lambda}}{2\sqrt{2}s}v^2\sin 2\beta+g^{'2}_1\tilde{Q}_S^2s^2+\Delta_{33}\,.
\label{62}
\end{equation}
In Eq.~(\ref{57})--(\ref{62}) $\Delta_{ij}$ denote the loop corrections.

If all SUSY breaking parameters as well as the VEV of the SM singlet field $S$ are considerably larger than the EW
scale, the mass matrix (\ref{56})--(\ref{62}) has a hierarchical structure. In the field basis $(h,\,H,\,N)$
all off--diagonal elements of this matrix are relatively small $\sim M_S M_Z$. Therefore the masses of the heaviest
CP--even Higgs bosons are closely approximated by the diagonal entries $M_{22}^2$ and $M_{33}^2$ which are expected
to be of the order of the SUSY breaking scale $M_S^2$. These two CP--even Higgs bosons are predominantly formed by
the components of the field basis $H$ and $N$. Because the minimal eigenvalue of a Hermitian matrix does not
exceed its smallest diagonal element the lightest Higgs state in the CP-even sector (approximately $h$) remains always light
irrespective of the SUSY breaking scale, i.e. $m^2_{h_1} < M_{11}^2$ like in the MSSM and NMSSM. In the interactions with
other SM particles this state manifests itself as a SM-like Higgs boson if $M_S \gg M_Z$.

As follows from Eqs.~(\ref{51}), (\ref{54}) and (\ref{56})--(\ref{62}) at the tree level the spectrum of the Higgs bosons
depends on four variables only:
\begin{equation}
\lambda\,,\qquad s\,,\qquad \tan\beta\,,\qquad A_{\lambda}\,.
\label{63}
\end{equation}

\section{Higgs spectrum}

The qualitative pattern of the Higgs spectrum in the E$_6$SSM inspired is determined by the Yukawa coupling $\lambda$.
Let us start our analysis here from the MSSM limit of the E$_6$SSM when $\lambda \ll g'_1$. In the case when $\lambda$ goes to zero
$s$ has to be sufficiently large so that $\mu = \lambda s/\sqrt{2}$ is held fixed in order to give an acceptable chargino mass and EWSB.
The diagonal entry $M_{33}^2$ that is set by the mass of the $Z'$ boson tends to be substantially larger than other elements of the
mass matrix (\ref{56})--(\ref{62}) in this scenario. From the first minimisation conditions (\ref{45}) one can see that such solution
can be obtained for very large and negative values of $m_S^2$. If $\mu\ll M_{Z'}$ and $m_A^2\ll M_{Z'}$ the CP--even Higgs mass matrix (\ref{56})--(\ref{62}) can be reduced to the block diagonal form by means of a small unitary transformation \cite{Miller:2003ay,Nevzorov:2004ge}
\begin{equation}
M'^2\approx\left(
\begin{array}{ccc}
M_{11}^2-\dfrac{M_{13}^4}{M_{33}^2} & M_{12}^2-\dfrac{M^2_{13}M^2_{32}}{M^2_{33}} & 0\\[0mm]
M_{21}^2-\dfrac{M_{23}^2 M_{31}^2}{M_{33}^2} & M_{22}^2-\dfrac{M_{23}^4}{M_{33}^2} & 0\\[0mm]
0 & 0 & M_{33}^2+\dfrac{M_{13}^4}{M_{33}^2}+\dfrac{M_{23}^4}{M_{33}^2}
\end{array}
\right)\,.
\label{64}
\end{equation}
For small values of $\lambda$ the top--left $2\times 2$ submatrix in Eq.~(\ref{64}) reproduces the mass matrix of the CP--even Higgs
sector in the MSSM. Such hierarchical structure of the mass matrix of the CP-even Higgs sector, implies that the
mass of the $Z'$ boson and the mass of the heaviest CP-even Higgs particle associated with $N$ are almost degenerate.
In other words the singlet dominated CP-even state is always very heavy and decouples from the rest of the spectrum, which makes
the Higgs spectrum indistinguishable from the one in the MSSM. Its mass is determined by the VEV of the SM singlet field and does not
change much if the other parameters $\lambda$, $\tan\beta$ and $A_{\lambda}$ ($m_A$) vary.
The masses of the second lightest Higgs scalar, that is predominantly $H$, the Higgs pseudoscalar and the charged Higgs states
grow when $m_A$ rises providing the degeneracy of the corresponding states at $m_A$ when $m_A$ is much larger than $M_Z$ but
is less than $M_{Z'}$. In this case the expression for the SM-like Higgs mass $m_{h_{1}}^2$ is essentially
the same as in the MSSM.

\begin{figure}
\centering
\includegraphics[width=7.7cm]{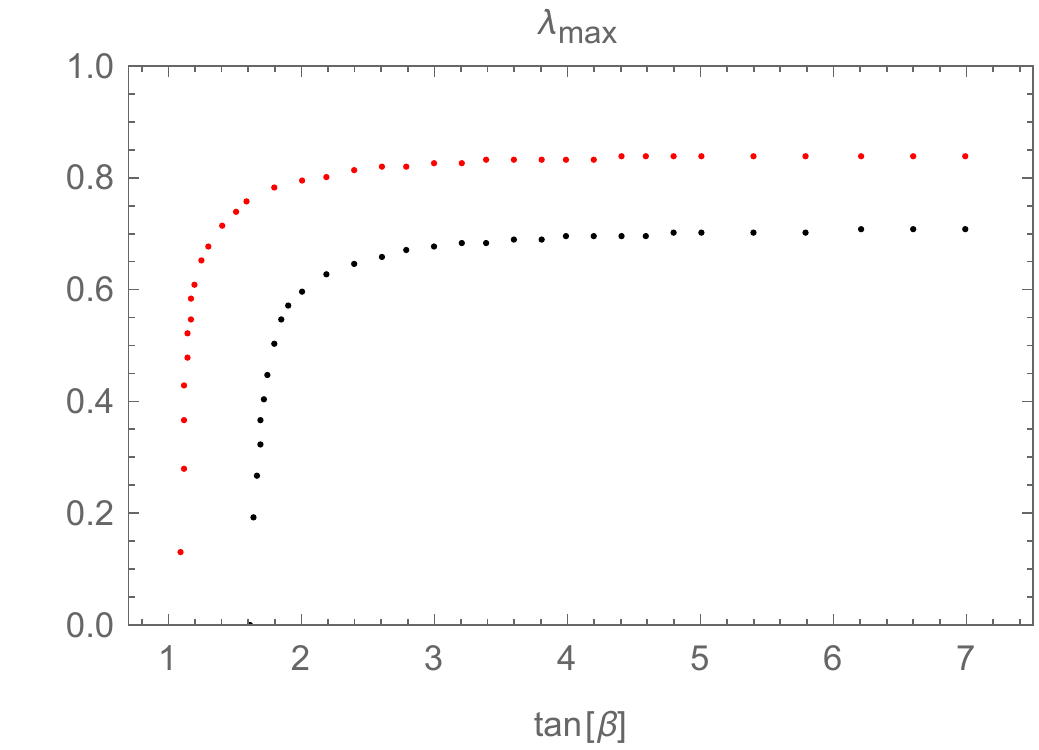}
\includegraphics[width=7.7cm]{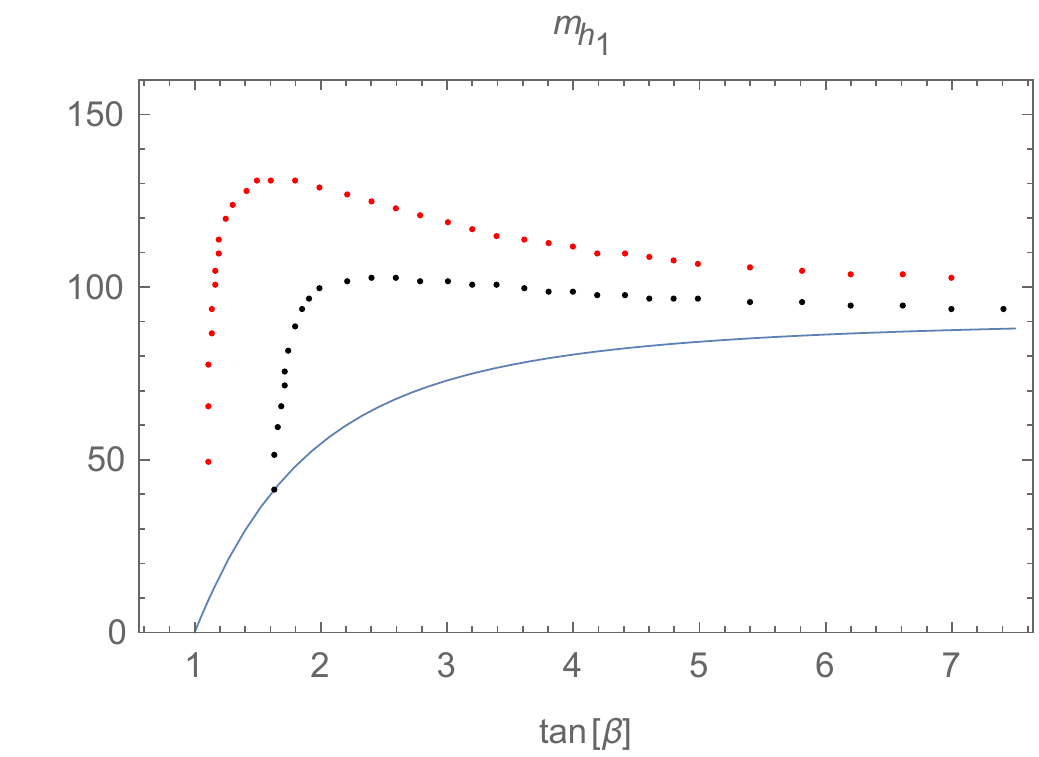}
\caption{
{\it Left:} Upper limit on $\lambda$ versus $\tan\beta$ in the NMSSM (lower dotted line)
and E$_6$SSM (upper dotted line). {\it Right:} Tree--level upper bound on the lightest Higgs boson
mass as a function of $\tan\beta$ in the MSSM (solid line), NMSSM (lower dotted line) and E$_6$SSM (upper dotted line).}
\label{essmfig2}
\end{figure}

When $\lambda\ge g'_1$ the qualitative pattern of the spectrum of the Higgs bosons is rather similar to the one
that arises in the NMSSM with the approximate PQ symmetry \cite{Miller:2003ay,Nevzorov:2004ge,Miller:2005qua,Miller:2003hm,King:2014xwa}.
In the NMSSM and E$_6$SSM the growth of the Yukawa coupling $\lambda$ at low energies entails the increase of its value at the GUT
scale $M_X$ resulting in the appearance of the Landau pole that spoils the applicability of perturbation theory at high energies \cite{Nevzorov:2001vj,Nevzorov:2002ub}. The requirement of validity of perturbation theory up to
the scale $M_X$ sets an upper limit on $\lambda(M_t)$ for each fixed value of $\tan\beta$ in these models.
In the E$_6$SSM the restrictions on the low energy values of $\lambda$ are weaker than in the NMSSM (see Figure~\ref{essmfig2}/left).
The presence of exotic matter change the running of the SM gauge couplings so that their values at the intermediate scale rise
when the number of extra $5+\overline{5}$--plets increases. In the RGEs that describe the evolution of the Yukawa couplings
within the NMSSM and E$_6$SSM the gauge couplings occur in the right--hand side of these equations with negative
sign. As a consequence the growth of the SM gauge couplings prevents the appearance of the Landau pole in the RG flow of the Yukawa
couplings. Therefore in the E$_6$SSM $\lambda(M_t)$ are allowed to be larger than in the NMSSM.
The upper bound on $\lambda(M_t)$ grows with increasing $\tan\beta$ since the top--quark Yukawa coupling decreases.
At large $\tan\beta$ this bound approaches the saturation limit. In the NMSSM and E$_6$SSM the maximal possible values of
$\lambda(M_t)$ are 0.71 and 0.84 respectively whereas the low energy value of $g'_1\approx g_1$ vary from $0.46$ to $0.48$.

If $\lambda\ge g'_1$ then $M_{22}^2$ tends to be the largest diagonal entry of the mass matrix (\ref{56})--(\ref{62}), i.e.
$M_{22}^2\gg M_{33}^2\gg M_{11}^2$. Relying on this mass hierarchy the approximate solutions for the Higgs masses
can be obtained. The perturbation theory method yields \cite{Kovalenko:1998dc,Nevzorov:2000uv,Nevzorov:2001um,Miller:2003ay,Nevzorov:2004ge}
\begin{equation}
m^2_{h_3}\approx M_{22}^2+\dfrac{M_{23}^4}{M_{22}^2}\,,\qquad
m^2_{h_2}\approx M_{33}^2-\dfrac{M_{23}^4}{M_{22}^2}+\dfrac{M_{13}^4}{M_{33}^2}\,,\qquad
m^2_{h_1}\approx M_{11}^2-\dfrac{M_{13}^4}{M_{33}^2}\,.
\label{65}
\end{equation}
In Eq.~(\ref{65}) all terms suppressed by inverse powers of $m_A^2$ or $M_{Z'}^2$, i.e.
$O(M_Z^4/m_A^2)$ and $O(M_Z^4/M_{Z'}^2)$, are neglected. At tree--level the masses of the Higgs bosons can written as
\begin{equation}
\begin{array}{rcl}
m^2_{h_3}&\approx & m^2_{H^{\pm}} \approx m_A^2\approx \dfrac{4\mu^2 x}{\sin^2 2\beta}\,,\qquad m^2_{h_2}\approx M_{Z'}^2\,,\\
m^2_{h_1}& \approx & \dfrac{\lambda^2}{2}v^2\sin^22\beta + M_Z^2\cos^22\beta +
g^{'2}_1 v^2\biggl(\tilde{Q}_{H_d}\cos^2\beta + \tilde{Q}_{H_u}\sin^2\beta\biggr)^2\\
&& - \dfrac{\lambda^4 v^2}{g^{'2}_1 \tilde{Q}_S^2}\biggl(1-x+\dfrac{g^{'2}_1}{\lambda^2}\biggl(\tilde{Q}_{H_d}\cos^2\beta
+\tilde{Q}_{H_u}\sin^2\beta\biggr)\tilde{Q}_S\biggr)^2\,,
\end{array}
\label{66}
\end{equation}
where $x=\dfrac{A_{\lambda}}{2\mu}\sin 2\beta$ and $\mu=\dfrac{\lambda}{\sqrt{2}}s$.
As evident from the explicit expression for $m^2_{h_1}$ given above at $\lambda^2\gg g^{2}_1$ the last term in this expression
dominates and the mass squared of the lightest Higgs boson tends to be negative if the auxiliary variable $x$ is not close to unity.
A negative eigenvalue of the mass matrix (\ref{57})--(\ref{62}) implies that the vacuum configuration (\ref{44}) ceases to be a minimum
and turns into a saddle point. Near this point there is a direction in field space along which the energy density decreases leading to
the instability of the vacuum configuration (\ref{44}). Thus large deviations of $x$ from unity pulls the mass squared of the lightest
Higgs boson below zero destabilising the vacuum. The requirement of stability of the physical vacuum therefore constrains
the variable $x$ around unity and limits the range of variations of $m_A$ from below and above. As a consequence the masses of the heaviest
CP--even, CP--odd and charged Higgs states are almost degenerate around $m_A$ and are confined in the vicinity of $\mu\, \tan\beta$.
They are considerably larger than the masses of the $Z'$ and lightest CP--even Higgs boson. Together with the experimental lower limit
on the mass of the $Z'$ boson it maintains the mass hierarchy in the spectrum of the Higgs particles \cite{King:2005jy}.

From the explicit analytic expression for $m^2_{h_1}$ it is apparent that at some value of $x$ (or $m_A$) the lightest CP--even Higgs boson
mass attains its maximum value. It corresponds to the value of $x$ for which the fourth term in the expression for $m^2_{h_1}$ vanishes.
In this case the mass squared of the lightest Higgs boson coincides with the theoretical upper bound on $m^2_{h_1}$ given by $M_{11}^2$.
The sum of the first and second terms in the expression for $M_{11}^2$ are similar to the tree--level upper bound on $m^2_{h_1}$ in the
NMSSM \cite{Durand:1988rg,Drees:1988fc}.  The third term in Eq.~(\ref{57}) is a contribution coming from the additional $U(1)_{N}$
$D$--term in the Higgs scalar potential (\ref{39})--(\ref{42}). At tree--level the upper bound on the lightest Higgs mass in the E$_6$SSM
depends on $\lambda$ and $\tan\beta$ only. Using the obtained theoretical restrictions on the low energy values of $\lambda$ as a function
of $\tan\beta$, one can compute the maximum possible value of $m_{h_1}$ for each particular choice of $\tan\beta$.

The tree--level upper bound on the mass of the lightest Higgs scalar in the E$_6$SSM is presented in Figure~\ref{essmfig2}
(see Figure~\ref{essmfig2}/right) and compared to the corresponding bounds in the
MSSM and NMSSM. At moderate values of $\tan\beta \sim 1-3$ the theoretical restriction on lightest Higgs boson mass in the
E$_6$SSM and NMSSM exceeds the corresponding limit in the MSSM because of the extra contribution to $M_{11}^2$ induced by the first term
in the right hand side of Eq.~(\ref{57}) which comes from the additional $F$--term in the Higgs scalar potentials of the E$_6$SSM and NMSSM.
For such values of $\tan\beta$ in the E$_6$SSM and NMSSM this contribution to the upper bound on $m_{h_1}$ dominates.
Its size is determined by the Yukawa coupling $\lambda$. Since the upper limit on the coupling $\lambda$ caused by the validity
of perturbation theory in the NMSSM is more stringent than in the E$_6$SSM the tree--level theoretical restriction on $m_{h_1}$
in the NMSSM is considerably less than in the E$_6$SSM at moderate values of $\tan\beta$. In the framework of the E$_6$SSM the upper bound
on $m_{h_1}$ attains a maximum value of $130\,\mbox{GeV}$ at $\tan\beta=1.5-1.8$. So large tree-level theoretical restriction on the mass
of the lightest Higgs scalar means that in this model the contribution of loop corrections to $m_{h_1}^2$ is not needed to be as big as
in the MSSM and NMSSM in order to get the SM-like Higgs boson with mass around $125\,\mbox{GeV}$.

With increasing $\tan\beta$ the contribution to $M_{11}^2$ associated with the first term in the right hand side of Eq.~(\ref{57})
falls quite rapidly and becomes negligibly small as $\tan\beta\gg 10$. In contrast the contribution of the $SU(2)_W$ and $U(1)_Y$ $D$--terms
to $M_{11}^2$ (second term in the right hand side of Eq.~(\ref{57})) grows when $\tan\beta$ increases. At $\tan\beta > 4$ it exceeds
$\dfrac{\lambda^2}{2} v^2 \sin^2 2\beta$ and gives the dominant contribution to the tree--level theoretical restriction on $m_{h_1}$.
Therefore with increasing $\tan\beta$ the upper bound on the lightest Higgs boson mass in the NMSSM diminishes and approaches
the corresponding limit in the MSSM. In the case of the E$_6$SSM the third term in the right hand side of Eq.~(\ref{57}),
that comes from the extra $U(1)_{N}$ $D$--term contribution to the Higgs scalar potential (\ref{39})--(\ref{42}),
gives the second largest contribution to $M_{11}^2$ at very large values of $\tan\beta$. Because of the contribution of this term
the tree--level theoretical restriction on the mass of the lightest Higgs scalar in the E$_6$SSM, which also diminishes when $\tan\beta$ rises,
is still $6-7\,\mbox{GeV}$ larger than the ones in the MSSM and NMSSM even at very large values of $\tan\beta$.
As a consequence at large $\tan\beta$ the presence of the 125-GeV Higgs boson in the particle spectrum of the E$_6$SSM does not require as
large contribution of loop corrections to $m_{h_1}^2$ as in the MSSM and NMSSM.

The inclusion of loop corrections substantially increases the mass of the lightest Higgs scalar in SUSY models.
The dominant contribution comes from the loops involving the top quark and its superpartners because of the large
top-quark Yukawa coupling $h_t$. Within the MSSM leading one--loop and two--loop corrections to $m_{h_1}$ increase the upper bound
on the lightest Higgs boson mass, which does not exceed Z-boson mass ($M_Z\simeq 91.2\,\mbox{GeV}$) at the tree--level
\cite{Flores:1982pr,Inoue:1982ej}, from $M_Z$ to $130\,\mbox{GeV}$ (see \cite{Djouadi:2005gj} and references therein).
In the leading approximation two--loop upper bound on the lightest Higgs boson mass in the E$_6$SSM can be presented in the following
form \cite{King:2005jy}
\begin{equation}
\begin{array}{c}
m_{h_1}^2\le\biggl[\dfrac{\lambda^2}{2}v^2\sin^22\beta+M_Z^2\cos^22\beta+g^{'2}_1v^2\biggl(\tilde{Q}_{H_d}\cos^2\beta+
\tilde{Q}_{H_u}\sin^2\beta\biggr)^2\biggr]\left(1-\dfrac{3h_t^2}{8\pi^2}l\right)\\
+\dfrac{3 m_t^4}{2\pi^2 v^2}\left\{\dfrac{1}{2} U_t+ l + \dfrac{1}{16\pi^2}\biggl(\dfrac{3}{2}h_t^2-8g_3^2\biggr)(U_t+l)l\right\}\,,\\
m_t(M_t)=\dfrac{h_t(M_t)}{\sqrt{2}}v\sin\beta\,,\qquad
U_t=2\dfrac{X_t^2}{M_S^2}\biggl(1-\dfrac{1}{12}\dfrac{X_t^2}{M_S^2}\biggr)\,,\qquad l=\ln\biggl[\dfrac{M_S^2}{m_t^2}\biggr]\,,
\end{array}
\label{67}
\end{equation}
where $X_t$ is a stop mixing parameter, $M_S$ is a SUSY breaking scale defined as $m_Q^2=m_U^2=M_S^2$ while $m_Q^2$ and $m_U^2$ are soft
scalar masses of superpartners of the left--handed and right--handed components of the $t$--quark respectively.
Here the value of $m_t(M_t)$ can be computed using the world average mass of the top quark $M_t=173.1\pm 0.9\,\mbox{GeV}$
(see \cite{Tanabashi:2018oca}) and the relationship between the $t$--quark pole ($M_t$) and running ($m_t(Q)$) masses \cite{mtMS-1,mtMS-2}
\begin{equation}
m_t(M_t)=M_t\biggl[1-1.333\dfrac{\alpha_s(M_t)}{\pi}-
9.125\left(\dfrac{\alpha_s(M_t)}{\pi}\right)^2\biggr]\,.
\label{68}
\end{equation}

Eq.(\ref{67}) is just a simple generalization of the approximate expressions for the theoretical restriction on the lightest Higgs boson mass
obtained in the MSSM \cite{Carena:1995wu} and NMSSM \cite{Ellwanger:1999ji}. At $\lambda=0$ and $g'_1=0$ the right--hand side of Eq.~(\ref{67})
coincides with the theoretical bound on the lightest Higgs mass in the MSSM. The analytic approximation of the two--loop effects given above
slightly underestimates the full two--loop corrections. In the MSSM the approximate expression (\ref{67}) results in the value of the lightest
Higgs mass which is typically a few $\mbox{GeV}$ lower than the one which is computed using the Suspect \cite{suspect} and FeynHiggs
\cite{feynhiggs-1,feynhiggs-2,feynhiggs-3,feynhiggs-4} packages. It was shown that in the two-loop approximation the mass of
the lightest Higgs scalar in the E$_6$SSM does not exceed $150\,\mbox{GeV}$ \cite{King:2005jy}.

Although the inclusion of loop corrections changes considerably the lightest Higgs boson mass in the E$_6$SSM, it does not change the
the qualitative pattern of the spectrum of the Higgs states for $\lambda \ll g'_1$ and $\lambda > g'_1$.
The mass of the SM singlet dominated CP-even state is always set by $M_{Z'}$ whereas another Higgs scalar, CP--odd and charged Higgs bosons
have masses close to $m_A$. In the phenomenologically viable scenarios the masses of all Higgs particles except the lightest Higgs state
are much larger than $M_Z$. Moreover when $\lambda > g'_1$ and, in particular, in the part of the E$_6$SSM parameter space where
the lightest Higgs boson can be heavier than $100-110\,\mbox{GeV}$ even at tree-level, the heaviest CP--even, CP--odd and charged Higgs states
lie beyond the multi-TeV range and therefore cannot be detected at the LHC experiments.

\section{LHC signatures}

We now turn to the LHC signatures of the E$_6$SSM, that permit to distinguish this SUSY model from the MSSM or NMSSM.
As discussed earlier, in the simplest phenomenologically viable scenarios the lightest exotic fermion $\tilde{H}^0_1$ should have mass
$m_{\tilde{H}^0_1}\ll 1\,\mbox{eV}$. At the same time next--to--lightest exotic fermion $\tilde{H}^0_2$ may be considerably heavier.
Let us assume that all sparticles and exotic states except $\tilde{H}^0_1$ and $\tilde{H}^0_2$ are
rather heavy and can be integrated out. In particular, the parameters are chosen so that all fermion components
of the supermultiplets $H^{u}_{\alpha}$ and $H^{d}_{\alpha}$ are heavier than $100\,\mbox{GeV}$, whereas $s\approx 12\,\mbox{TeV}$.
In this limit the part of the Lagrangian, that describes the interactions of $\tilde{H}^0_1$ and $\tilde{H}^0_2$
with the $Z$ boson and the SM-like Higgs particle can be presented in the following form:
\begin{equation}
\begin{array}{c}
\mathcal{L}_{Zh}=\sum_{\alpha,\beta}\dfrac{M_Z}{2 v}Z_{\mu}
\biggl(\tilde{H}^{0T}_{\alpha}\gamma_{\mu}\gamma_{5}\tilde{H}^0_{\beta}\biggr) R_{Z\alpha\beta} +
\sum_{\alpha,\beta} X^{h}_{\alpha\beta} \biggl(\tilde{H}^{0T}_{\alpha} \tilde{H}^0_{\beta}\biggr) h\,,
\end{array}
\label{69}
\end{equation}
where $\alpha,\beta=1,2$. Although $\tilde{H}^0_1$ and $\tilde{H}^0_2$ are substantially lighter than $100\,\mbox{GeV}$, their couplings
to the $Z$ boson and other SM particles can be negligibly small because these states are predominantly the fermion components of the
superfields $S_{\alpha}$. Therefore any possible signal, which $\tilde{H}^0_1$ and $\tilde{H}^0_2$ could give
rise to at former and present collider experiments, would be extremely suppressed and such states could remain undetected.

The couplings of the SM-like Higgs boson $h_1$ to $\tilde{H}^0_1$ and $\tilde{H}^0_2$ are determined by the masses of these lightest
exotic states \cite{Hall:2010ix}. Since $\tilde{H}^0_1$ is extremely light, it does not affect Higgs phenomenology.
The absolute value of the coupling of $h_1$ to the second lightest exotic particle $|X^{h}_{22}|\simeq |m_{\tilde{H}^0_2}|/v$ \cite{Hall:2010ix}.
This coupling gives rise to the decays of $h_1$ into $\tilde{H}^0_2$ pairs with partial width given by
\begin{equation}
\Gamma(h_1\to\tilde{H}^0_{2}\tilde{H}^0_{2})=\dfrac{|X^{h}_{22}|^2 m_{h_1}}{4\pi}
\biggl(1-4\dfrac{|m_{\tilde{H}^0_{2}}|^2}{m^2_{h_1}}\biggr)^{3/2}\,.
\label{70}
\end{equation}
The partial decay width (\ref{70}) depends rather strongly on $m_{\tilde{H}^0_2}$. To avoid the suppression of the branching
ratios for the lightest Higgs decays into SM particles we restrict our consideration here to the GeV-scale masses of the second
lightest exotic particle.

In order to compare the partial widths associated with the exotic decays of $h_1$ (\ref{70}) with the SM-like Higgs decay
rates into the SM particles a set of benchmark points (see Table~\ref{tab2}) is specified. In Table~\ref{tab2} the masses of
the heavy Higgs states are computed in the leading one-loop approximation. In the case of the lightest Higgs boson mass
the leading two-loop corrections are taken into account. In all benchmark scenarios the structure of the Higgs spectrum is
very hierarchical, the partial widths of the decays of $h_1$ into the SM particles are basically the same as in the SM.
Therefore in our analysis we use the results presented in \cite{King:2012is} where the corresponding decay rates
were computed within the SM for different values of the Higgs mass. When $m_{h_1}\simeq 125\,\mbox{GeV}$, the SM-like Higgs
state decays predominantly into $b$ quark. The corresponding branching ratio is about $60\%$ while the branching ratios
associated with Higgs decays into $WW$ and $ZZ$ are about $20\%$ and $2\%$, respectively \cite{King:2012is}.
The total decay width of such Higgs boson is about $4\,\mbox{MeV}$.

The benchmark scenarios (i)--(iv) presented in Table~\ref{tab2} demonstrate that the branching ratio of the exotic decays of $h_1$
changes from $0.2\%$ to $20\%$ when $m_{\tilde{H}^0_{2}}$ varies from $0.3\,\mbox{GeV}$ to $2.7\,\mbox{GeV}$ \cite{Nevzorov:2013tta}.
For smaller (larger) values of $m_{\tilde{H}^0_{2}}$ the branching ratio of these decays is even smaller (larger).
On the other hand, the couplings of $\tilde{H}^0_1$ and $\tilde{H}^0_2$ to the $Z$ boson are so small that these exotic fermions
could not be observed before. In particular, their contribution to the $Z$-boson width tend to be rather small.
After being produced $\tilde{H}^0_2$ sequentially decay into $\tilde{H}^0_1$ and fermion--antifermion pair via virtual $Z$.
Thus the exotic decays of $h_1$ result in two fermion--antifermion pairs and missing energy in the final state.
Nevertheless, since $|R_{Z12}|$ is quite small, $\tilde{H}^0_2$ tends to live longer than $10^{-8}\,\mbox{s}$
and typically decays outside the detectors. As a consequence, the decay channel $h_1\to\tilde{H}^0_2\tilde{H}^0_2$ normally gives
rise to an invisible branching ratio of the SM-like Higgs boson. Such invisible decays of $h_1$ take place in the benchmark
scenarios (i), (iii), and (iv). In the case of benchmark scenario (ii) $|R_{Z12}|$ is larger so that
$\tau_{\tilde{H}^0_{2}}\sim 10^{-11}\,\mbox{s}$ and some of the decay products of $\tilde{H}^0_2$ might be observed at the LHC.

\begin{table}[ht]
\centering
\begin{tabular}{|c|c|c|c|c|}
\hline
                              & i       &   ii      &   iii     &   iv    \\
\hline
$\lambda_{22}$                & -0.03   &   -0.012  &   -0.06   &   0  \\
$\lambda_{21}$                & 0       &   0       &   0       &   0.02   \\
$\lambda_{12}$                & 0       &   0       &   0       &   0.02   \\
$\lambda_{11}$                & 0.03    &   0.012   &   0.06    &   0  \\
$f_{22}$                      & -0.1    &   -0.1    &   -0.1    &   0.6    \\
$f_{21}$                      & -0.1    &   -0.1    &   -0.1    &   0.00245 \\
$f_{12}$                      & 0.00001 &   0.00001 &   0.00001 &   0.00245 \\
$f_{11}$                      & 0.1     &   0.1     &   0.1     &   0.00001 \\
$\tilde{f}_{22}$              & 0.1     &   0.1     &   0.1     &   0.6     \\
$\tilde{f}_{21}$              & 0.1     &   0.1     &   0.1     &   0.002 \\
$\tilde{f}_{12}$              & 0.000011&   0.000011&   0.000011&   0.002  \\
$\tilde{f}_{11}$              & 0.1     &   0.1     &   0.1     &   0.00001 \\
$|m_{\tilde{H}^0_1}|$/GeV  &$2.7\times 10^{-11}$  &   $6.5\times 10^{-11}$ & $1.4\times 10^{-11}$   & $0.31\times 10^{-9}$ \\
$|m_{\tilde{H}^0_2}|$/GeV  & 1.09    &   2.67    &   0.55    &   0.319   \\
$|R_{Z11}|$                   & 0.0036  &   0.0212  &   0.00090 &   $1.5\times 10^{-7}$ \\
$|R_{Z12}|$                   & 0.0046  &   0.0271  &   0.00116 &   $1.7\times 10^{-4}$ \\
$|R_{Z22}|$                   & 0.0018  &   0.0103  &   0.00045 &   0.106   \\
$X^{h_1}_{22}$                & 0.0044  &   0.0106  &   0.0022  &   0.00094 \\
$\mathrm{Br}(h\rightarrow \tilde{H}^0_2 \tilde{H}^0_2)$& 4.7\%   & 21.9\%              & 1.23\% & 0.22\% \\
$\mathrm{Br}(h\rightarrow b\bar{b})$                         & 56.6\%  & 46.4\%              & 58.7\% & 59.3\% \\
$\Gamma(h\rightarrow \tilde{H}^0_2 \tilde{H}^0_2)$/MeV & 0.194   & 1.106               & 0.049  & 0.0088 \\
\hline
\end{tabular}
\caption{Benchmark scenarios for $m_{h_1}\approx 125\,\mbox{GeV}$; the branching ratios and decay widths of the
lightest Higgs boson, the masses and couplings of $\tilde{H}^0_1$ and $\tilde{H}^0_2$ are calculated for
$s=12\,\mbox{TeV}$, $\lambda=0.6$, $\tan\beta=1.5$, $m_{H^{\pm}}\simeq m_{A}\simeq m_{h_3}\simeq 9497\,\mbox{GeV}$,
$m_{h_2}\simeq M_{Z'}\simeq 4450\,\mbox{GeV}$, $m_Q=m_U=M_S=4000\,\mbox{GeV}$ and $X_t=\sqrt{6} M_S$.}
\label{tab2}
\end{table}

Because $R_{Z12}$ is relatively small, $\tilde{H}^0_2$ may decay during or after Big Bang Nucleosynthesis (BBN) destroying
the agreement between the predicted and observed light element abundances. To preserve the success of the BBN, $\tilde{H}^0_2$ should
decay before BBN, i.e. its lifetime $\tau_{\tilde{H}^0_{2}}$ should not be longer than $1\,\mbox{s}$.
This requirement constrains $|R_{Z12}|$. Indeed, for $m_{\tilde{H}^0_{2}}=1\,\mbox{GeV}$
the absolute value of the coupling $R_{Z12}$ has to be larger than $1\times 10^{-6}$ \cite{King:2012wg}.
The constraint on $|R_{Z12}|$ becomes more stringent with decreasing $m_{\tilde{H}^0_{2}}$ because
$\tau_{\tilde{H}^0_{2}}\sim 1/(|R_{Z12}|^2 m_{\tilde{H}^0_{2}}^5)$. The results of our analysis indicate that
it is somewhat problematic to ensure that $\tau_{\tilde{H}^0_{2}}\le 1\,\mbox{s}$ if $m_{\tilde{H}^0_{2}}\le 100\,\mbox{MeV}$.

The presence of a $Z'$ gauge boson and exotic multiplets of matter that compose three $5+5^{*}$ representations of $SU(5)$
is another very peculiar feature of the E$_6$SSM. LHC signatures associated with these states are determined by the structure
of the particle spectrum that varies substantially depending on the choice of the parameters. At tree--level the masses of
the $Z'$ boson and fermion components of $5+5^{*}$ supermultiplets are set by the VEV of the SM singlet field $S$, that remains
a free parameter in this models. Therefore the masses of these states cannot be predicted. The lower experimental limits on the $Z'$ mass,
that comes from the direct searches $(pp\to Z'\to l^{+}l^{-})$ conducted at the LHC experiments, are already very stringent
and vary around $3.8-3.9\,\mbox{TeV}$ \cite{Aaboud:2017buh,Sirunyan:2018exx}. This means that the scenarios with $s<10-10.5\,\mbox{TeV}$
have been excluded. Possible $Z'$ decay channels in $E_6$ inspired SUSY models were studied in \cite{Accomando:2010fz,Gherghetta:1996yr}.

Assuming that $f_{\alpha\beta}$ and $\tilde{f}_{\alpha\beta}$ are very small the masses of the fermion components of extra $5+5^{*}$
supermultiplets of matter are given by
\begin{equation}
\mu_{D_i}=\dfrac{\kappa_i}{\sqrt{2}}\,s\,, \qquad\qquad
\mu_{H_{\alpha}}=\dfrac{\lambda_{\alpha}}{\sqrt{2}}\,s\,,
\label{71}
\end{equation}
where $\mu_{D_i}$ are the masses of the $SU(3)_C$ colour triplets of exotic quarks with electric charges $\pm 1/3$ and
$\mu_{H_{\alpha}}$ are the masses of the $SU(2)_W$ doublets of the Inert Higgsino states. Here we set $\kappa_{ij}=\kappa_i\delta_{ij}$
and $\lambda_{\alpha\beta}=\lambda_{\alpha}\delta_{\alpha\beta}$. The requirement of the validity of perturbation theory up to the GUT scale $M_X$
sets stringent upper bounds on the low--energy values of the Yukawa couplings $\kappa_i$ and $\lambda_{\alpha}$. Nevertheless the low--energy values
of these couplings are allowed to be as large as $g^{'}_1(q)\approx g_1(q)\approx 0.46-0.48$. On the other hand couplings $\kappa_i$ and
$\lambda_{\alpha}$ must be large enough to ensure that the exotic fermions are sufficiently heavy to avoiding conflict with direct
particle searches at present and former accelerators. Although nowadays there are clear indications that $Z'$ boson and sparticles
have to be rather heavy some of the exotic fermions can be relatively light in the E$_6$SSM. This happens, for example, if the Yukawa couplings
of the exotic particles $\kappa_{ij}$ and $\lambda_{\alpha}$ have hierarchical structure similar to the one observed in the ordinary quark and
lepton sectors. Then $Z'$ boson can be much heavier than $10\,\mbox{TeV}$ and the only manifestation of this SUSY extension of the SM
may be the presence of light exotic quark and/or Inert Higgsino states in the particle spectrum.

If the relatively light exotic quarks of the nature described above do exist, they might be accessed through direct pair hadroproduction.
The lifetime and decay modes of the lightest exotic quarks are determined by the operators that break the $Z_2^{H}$ symmetry.
Since in order to suppress FCNCs the Yukawa couplings of exotic particles to the quarks and leptons of the first two generations must be
rather small, here we assume that exotic states couple most strongly with the third family fermions and bosons.
Then, because the lightest exotic quarks are $R$--parity odd states, they decay either via
\begin{equation}
\overline{D} \to t + b + E^{\rm miss}_{T}+X\,,
\label{72}
\end{equation}
if exotic quarks $\overline{D}_i$ are diquarks or via
\begin{equation}
D \to t + \tau + E^{\rm miss}_{T} + X\,,\qquad\qquad D \to b + \nu_{\tau} + E^{\rm miss}_{T} + X\,,
\label{73}
\end{equation}
if exotic quarks of type $D$ are leptoquarks. Thus the pair production of light $D$-fermions at the LHC should result
in some enhancement of the cross sections of either $pp\to t\overline{t}b\overline{b}+E^{\rm miss}_{T}+X$ if exotic
quarks are diquarks or $pp\to t\overline{t}\tau^+{\tau^-}+E^{\rm miss}_{T}+X$ and $pp\to b\overline{b}+E^{\rm miss}_{T}+X$
if new quark states are leptoquarks.

In general exotic squarks tend to be considerably heavier than the exotic quarks because their masses are determined
by the soft SUSY breaking terms. Nevertheless the exotic squark associated with the heavy exotic quark can be relatively
light. This happens when the large mass of the heaviest exotic quark in the E$_6$SSM gives rise to the large mixing
in the corresponding exotic squark sector. Such mixing may result in the large mass splitting between the appropriate
mass eigenstates. As a consequence the lightest exotic squark may be much lighter than all other scalars. Moreover,
in principle, it can be even lighter than the lightest exotic quark. If this is a case then in the variants of the
E$_6$SSM with approximate $Z_2^{H}$ symmetry the lightest exotic squark decays into either
\begin{equation}
\tilde{D} \to t + b + +X\,,
\label{74}
\end{equation}
if it is a scalar diquark or
\begin{equation}
D \to t + \tau  + X\,,\qquad\qquad D \to b + \nu_{\tau} + X\,,
\label{75}
\end{equation}
if exotic squark is a scalar leptoquark. In the limit, when the couplings of this sfermion to the quarks and leptons of
the first two generations are rather small, the lightest exotic squarks can only be pair produced at the LHC.
Therefore the presence of light $\tilde{D}$ in the particle spectrum is expected to lead to some enhancement of
the cross sections of either $pp\to t\overline{t}b\overline{b}+X$ if exotic squarks are diquarks
or $pp\to t\overline{t}\tau^+{\tau^-}+X$ and $pp\to b\overline{b}+E^{\rm miss}_{T}+X$ if these squarks are leptoquarks.
On the other hand in the variants of the E$_6$SSM with exact $\tilde{Z}_2^{H}$ symmetry
the $Z_{2}^{E}$ symmetry conservation implies that the final state in the decay of $\tilde{D}$ should always contain the lightest exotic
fermion $\tilde{H}^0_1$ \cite{Nevzorov:2012hs}. Because the lightest exotic squark is $R$--parity even state whereas $\tilde{H}^0_1$
is $R$--parity odd particle the final state in the decay of $\tilde{D}$ should also involve the lightest ordinary neutralino to ensure
that $R$--parity is conserved. As a consequence in such models the decay patterns of the lightest exotic squarks and their
LHC signatures are rather similar to the ones that appear in the case of the lightest exotic quarks. The presence of relatively light
exotic quark and squark can substantially modify the LHC signatures associated with the gluinos \cite{Belyaev:2012si}.

Several experiments at LEP, HERA, Tevatron and LHC have searched for colored objects that decay into either a pair of quarks or
quark and lepton. Most searches focus on leptoquarks or diquarks which have integer--spin so that they can be either scalars or vectors.
Such objects can be coupled directly to either a pair of quarks or to quark and lepton. The most stringent constraints on the masses of
scalar leptoquarks and scalar diquarks come from the non-observation of these exotic states at the LHC experiments.
ATLAS and CMS collaborations ruled out first and second generation scalar leptoquarks (i.e. leptoquarks that couple to the first and
second generation fermions respectively) that have masses below $1230-1560\,\mbox{GeV}$ depending on the branching ratios of their decays
\cite{Sirunyan:2018btu,Aaboud:2019jcc,Sirunyan:2018ryt}. The experimental limits on the masses of the third generation scalar
leptoquarks are somewhat weaker. ATLAS and CMS collaborations excluded such exotic objects if they have masses below $800-1000\,\mbox{GeV}$
\cite{Sirunyan:2018ruf,Sirunyan:2018nkj,Aaboud:2019bye}. The experimental lower limits on the masses of dijet resonances
including diquarks tend to be considerably higher \cite{Sirunyan:2018xlo}.

\begin{figure}
\centering
\includegraphics[width=12cm,angle=90]{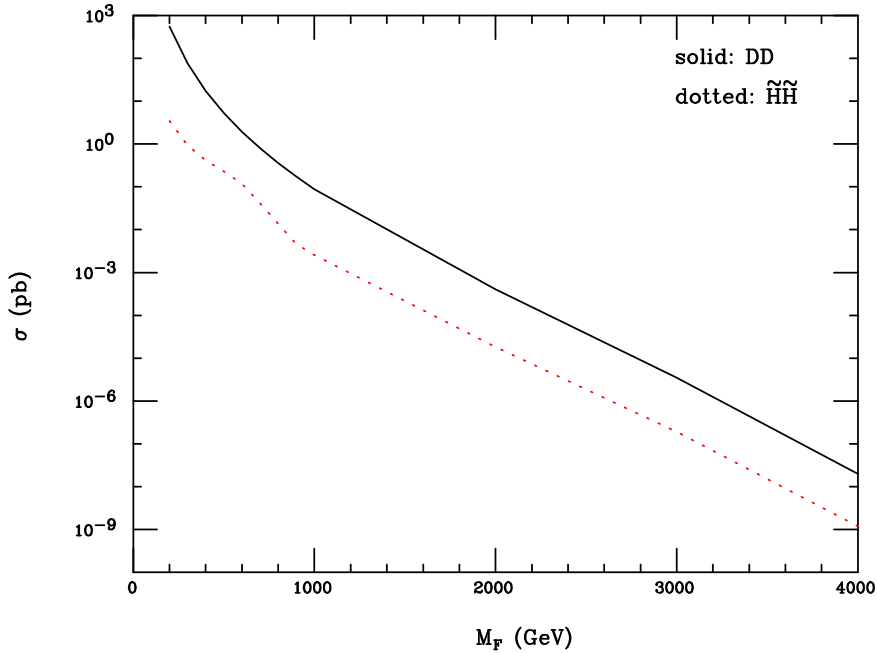}
\caption{Cross section at the LHC for pair production of exotic $D$--quarks (via QCD interactions)
as well as $SU(2)_W$ doublets of the Inert Higgsinos $\tilde{H}$ (via EW interactions), as a function of their (common) mass, denoted by $M_F$.}
\label{essmfig3}
\end{figure}

However the LHC lower bounds on the masses of exotic quarks/squarks are not always directly applicable in the case of the E$_6$SSM.
For instance, it is expected that scalar diquarks are mostly produced singly at the LHC and decay into final state that contains two quarks.
At the same time within the E$_6$SSM the couplings of all exotic scalars to the fermions of the first and second generation should be rather
small to avoid processes with non--diagonal flavour transitions. Therefore in this SUSY model diquarks can only be pair produced.
It is also worthwhile to point out that the lightest exotic quarks in the E$_6$SSM give rise to collider signatues which are very different
from the commonly established ones associated with the scalar leptoquarks or diquarks that have been thoroughly studied. Indeed, it is commonly
assumed that these scalars decay into quark--quark or quark--lepton without missing energy. On the other hand in the E$_6$SSM exotic quarks are
fermions and therefore $R$--parity odd states. Thus $R$--parity conservation necessarily leads to the missing energy and transverse momentum
in the final state. Because of this the pair production of the lightest exotic quark with the baryon number, which is twice larger than that
of ordinary ones, and the pair production of gluinos at the LHC may result in the enhancement of the same cross section of
$pp\to t\bar{t}b\bar{b}+E^{\rm miss}_{T}+X$.

The $SU(2)_W$ doublets of the Inert Higgsino states can be also light or heavy depending on their free parameters.
When at least one coupling $\lambda_{\alpha}$ is of the order of unity it can induce a large mixing in the Inert Higgs
sector that may lead to relatively light Inert Higgs bosons. Since these bosons have very small couplings to the fermions of the first
and second generation at the LHC the corresponding states can be produced in pairs via off--shell $W$ and $Z$--bosons.
As a consequence their production cross section is relatively small even when these particles have masses below the TeV scale.
After being produced they sequentially decay into the third generation fermions that should lead to some enlargement of the
cross sections of $pp\to Q\bar{Q}Q^{'}\bar{Q}^{'}$ and $pp\to Q\bar{Q}\tau^+{\tau^-}$ production, where $Q$ and $Q'$ are
heavy quark of the third generation.

As follows from Eq.~(\ref{71}) the lightest Inert Higgsinos can be relatively light if the corresponding Yukawa coupling
$\lambda_{\alpha}$ is sufficiently small. If all other exotic states and sparticles are rather heavy the corresponding
fermionic states can be produced at the LHC via weak interactions only. As a consequence their production cross section is
considerably smaller than the production cross section of the exotic quarks (see Figure~\ref{essmfig3}).
The Inert Higgsino states decay predominantly into the lightest exotic fermions ($\tilde{H}^0_1$ or $\tilde{H}^0_2$)
as well as an on-shell $Z$ or $W$ boson. Thus when pair produced Inert Higgsinos decay
they should lead to some enhancements in the rates of $pp \to ZZ+ E^{\rm miss}_{T} + X$, $pp \to WZ + E^{\rm miss}_{T} + X$
and $pp \to WW + E^{\rm miss}_{T} + X$. Similar enhancement of these cross sections could be caused
by the pair production of ordinary chargino and neutralino in the MSSM if the mass of the LSP is negligibly small.
Using the corresponding results of the analysis of ATLAS and CMS collaborations \cite{Aaboud:2018jiw,Aaboud:2018sua,Sirunyan:2018ubx}
one can conclude that the mass of the $SU(2)_W$ doublets of the Inert Higgsino states has to be larger than $650\,\mbox{GeV}$.

\section{Conclusions}

The breakdown of an extended gauge symmetry in the string-inspired $E_6$ GUTs may result in a variety
of extensions of the SM with softly broken SUSY at low energies including models based on the SM gauge group,
like the MSSM and NMSSM, as well as $U(1)$ extensions of the MSSM, etc. Among $U(1)$ extensions of the MSSM
inspired by $E_6$ GUTs there is unique choice of Abelian $U(1)_{N}$ gauge symmetry that allows zero
charges for right-handed neutrinos and thus a high scale see-saw mechanism. In the $U(1)_{N}$ extension
of the MSSM the lepton asymmetry, which may be induced by the heavy right-handed neutrino decays, can be
partially converted into baryon asymmetry via sphaleron processes \cite{Kuzmin:1985mm,Rubakov:1996}.
In this Exceptional Supersymmetric Standard Model (E$_6$SSM) the extra $U(1)_{N}$ gauge symmetry forbids
the term $\mu H_d H_u$ in the superpotential, but permits the term $\lambda S(H_u H_d)$, where $S$ is a
SM singlet superfield that carries $U(1)_{N}$ charge. When $S$ develops VEV breaking $U(1)_{N}$ gauge symmetry
it also gives rise to an effective $\mu$ term. Thus within the E$_6$SSM the $\mu$ problem of the MSSM is solved
in a similar way to that in the NMSSM, but without the accompanying problems of singlet tadpoles or domain walls.

In this review article we discussed the particle content, the global symmetries, which allows to suppress
FCNCs and rapid proton decay, as well as the RG flow of gauge couplings in the E$_6$SSM.
The low energy matter content of this SUSY model includes three copies of $27_i$ representations of $E_6$
so that anomalies get canceled generation by generation. In addition an extra pair of $SU(2)_W$ doublets
$L_4$ and $\overline{L}_4$ should survive to low energies to ensure high energy gauge coupling unification.
As a consequence the E$_6$SSM involves extra matter beyond the MSSM contained in three supermultiplets of
exotic charge 1/3 quarks ($D_i$ and $\overline{D}_i$), two pairs of $SU(2)_W$ doublets of Inert Higgs states,
three SM singlet superfields which carry $U(1)_{N}$ charges, $L_4$, $\overline{L}_4$ and $Z'$ vector superfield.
As in the MSSM, the gauge symmetry of the E$_6$SSM does not forbid baryon and lepton number violating interactions
that give rise to rapid proton decay. Moreover in general relatively light exotic states induce unacceptably large
flavor changing processes. To suppress the most dangerous baryon and lepton number violating operators
one can impose either $Z_2^L$ or $Z_2^B$ discrete symmetry which implies that the exotic quarks are
either diquarks (Model I) or leptoquarks (Model II). In order to avoid the appearance of the FCNCs
at the tree level one can postulate an approximate $Z^{H}_2$ symmetry, under which all supermultiplets of matter
except a pair of Higgs doublets ($H_{d}$ and $H_{u}$) and one SM singlet superfield $S$ are odd.
Instead of $Z^{H}_2$, $Z_2^L$ and $Z_2^B$ one can use a single discrete $\tilde{Z}^{H}_2$ symmetry which forbids
operators giving rise to too rapid proton decay and tree-level flavor-changing transitions. The Higgs supermultiplets
$H_u$, $H_d$ and $S$ as well as $L_4$ and $\overline{L}_4$ are even under the $\tilde{Z}^{H}_2$ symmetry
whereas all other matter fields are odd. In this case the exotic quarks are leptoquarks.

The results of the analysis of the two--loop RG flow within the E$_6$SSM were presented taking into account
kinetic term mixing between $U(1)_Y$ and $U(1)_{N}$ factors. If there is no mixing between $U(1)_Y$ and $U(1)_{N}$
near the GUT scale $M_X$ then the off--diagonal gauge coupling, which describes such mixing, remains negligibly small
at any intermediate scale between $M_X$ and TeV scale. In this limit the gauge coupling of the extra $U(1)_{N}$ is
always close to the $U(1)_{Y}$ gauge coupling. On the other hand the values of the SM gauge couplings at high energies
are considerably larger in the E$_6$SSM than in the MSSM due to the presence of the extra supermultiplets of exotic
matter in the $U(1)_{N}$ extensions of the MSSM. Our analysis revealed that the gauge coupling unification in the E$_6$SSM
can be achieved for phenomenologically acceptable values of $\alpha_3(M_Z)$, consistent with the central measured
low energy value of this coupling.

Because of the larger gauge couplings the theoretical restrictions on the low energy values of the Yukawa couplings coming
from the requirement of the validity of perturbation theory up to the scale $M_X$ get relaxed in the E$_6$SSM as compared
with the MSSM and NMSSM. As a consequence for moderate values of $\tan\beta$ the tree--level upper bound on the SM--like
Higgs boson mass can be considerably bigger in the E$_6$SSM than in the MSSM and NMSSM. In this SUSY model it
can be bigger than $115-125\,\mbox{GeV}$ so that the contribution of loop corrections to the mass of the lightest
Higgs scalar is not needed to be as large as in the MSSM and NMSSM in order to obtain $125\,\mbox{GeV}$ Higgs boson.

In this article the gauge symmetry breaking and the spectrum of the Higgs bosons within the E$_6$SSM were reviewed as well.
In the $U(1)_{N}$ extensions of the MSSM the SM singlet Higgs field $S$ is required to acquire a very large VEV
$\langle S\rangle=s/\sqrt{2}$, where $s>10\,\mbox{TeV}$, to ensure that the $Z'$ boson and exotic fermions gain
sufficiently large masses. In particular, the results of the analysis of the LHC data imply that the $U(1)_{N}$ gauge boson
must be heavier than $3.8-3.9\,\mbox{TeV}$. When CP--invariance is preserved, the E$_6$SSM Higgs spectrum includes
three CP--even, one CP--odd and two charged bosons. The SM singlet dominated CP--even state is almost degenerate with
the $Z'$ gauge boson. The masses of another CP--even and charged Higgs bosons are set by the mass of Higgs pseudoscalar
$m_A$. All these states tend to be substantially heavier than the lightest Higgs scalar that manifests itself in the interactions
with other SM particles as a SM-like Higgs boson. In the part of the E$_6$SSM parameter space, where the lightest Higgs state
can be heavier than $100-110\,\mbox{GeV}$ at the tree--level, all other Higgs bosons lie beyond the multi-TeV range and
therefore cannot be discovered at the LHC.

We also considered possible manifestations of the E$_6$SSM that may be observed at the LHC in the near future.
The simplest phenomenologically viable scenarios imply that LSP and NLSP are the lightest exotic states ($\tilde{H}^0_1$ and
$\tilde{H}^0_2$), which are formed by the fermion components of the SM singlet superfields $S_{\alpha}$. One of these
fermions $\tilde{H}^0_1$ should be much lighter than $1\,\mbox{eV}$ composing hot dark matter in the Universe. Such states give
only a very minor contribution to the dark matter density. The NLSP $\tilde{H}^0_2$ can have mass of the order of $1\,\mbox{GeV}$
giving rise to nonstandard decays of the $125\,\mbox{GeV}$ Higgs boson. Since $\tilde{H}^0_2$ tends to live longer than $10^{-8}\,\mbox{sec}.$
it decays outside the detectors. Therefore the decay channel $h_1\to\tilde{H}^0_2\tilde{H}^0_2$ results in an invisible branching ratio
of the SM-like Higgs state. The corresponding branching ratio can be as large as 20\%.

Other possible manifestations of the E$_6$SSM, which can permit to distinguish this model from the MSSM or NMSSM, are associated with the
presence of the $Z'$ gauge boson and exotic supermultiplets of matter that compose three $5+5^{*}$ representations of $SU(5)$.
The most spectacular LHC signals can come from the exotic color states and $Z'$. The production of the $Z'$ boson should lead to unmistakable
signal $pp\to Z'\to l^{+}l^{-}$ at the LHC. Assuming that the $Z_2^H$ symmetry is mainly broken by the operators involving quarks and leptons
of the third generation, the pair production of the lightest exotic quarks with masses in a few TeV range can give rise to the enhancement
of the cross section of either $pp\to t\overline{t}b\overline{b}+E^{\rm miss}_{T}+X$ if exotic quarks are diquarks or
$pp\to t\overline{t}\tau^+{\tau^-}+E^{\rm miss}_{T}+X$ and $pp\to b\overline{b}+E^{\rm miss}_{T}+X$ if exotic quarks are leptoquarks.
Because of the large mass splitting in the exotic squark sector, which can be caused by the heavy $D$-fermion, one of the exotic squarks
can be relatively light. If this is a case then the pair production of the superpartners of $D$-fermions may result in some enlargement
of the cross sections of either $pp\to t\overline{t}b\overline{b}+X$ when exotic squarks are diquarks or $pp\to t\overline{t}\tau^+{\tau^-}+X$
and $pp\to b\overline{b}+E^{\rm miss}_{T}+X$ if these squarks are leptoquarks. As compared with the exotic quarks and squarks
the production of Inert Higgs bosons and Inert Higgsinos is rather suppressed at the LHC.
The discovery of $Z'$ and new exotic states predicted by the E$_6$SSM would point towards an underlying $E_6$ gauge structure at high
energies and open a new era in elementary particle physics.

\vspace{0.5cm}
\section*{Acknowledgements}
R.N. would like to thank P.~Athron, E.~Boos, M.~Binjonaid, S.~Demidov, M.~Dubinin, D.~Gorbunov, D.~Harries, M.~Libanov, D.~Kazakov,
M.~M\"{u}hlleitner, V.~Rubakov, M.~Sher, A.~W.~Thomas, S.~Troitsky, X.~Tata and A.G.~Williams for fruitful discussions.
The work of S. F. K and S. M. is supported by the Science and Technology Facilities Council, Consolidated Grant number ST/L000296/1.
S. F. K. also acknowledges the European Union's Horizon 2020 Research and Innovation programme under
Marie Sk\l{}odowska-Curie grant agreements Elusives ITN No.~674896 and InvisiblesPlus RISE No.~690575.
S.M. further acknowledges partial financial support through the NExT Institute.


\begin{thebibliography}{999}
\bibitem{Georgi:1974sy}
Georgi, H.; Glashow, S. L. Unity Of All Elementary Particle Forces. {\em Phys. Rev. Lett.} {\bf 1974}, {\em 32}, 438.
\bibitem{Minkowski:1977sc}
Minkowski, P. $\mu \to e\gamma$ at a Rate of One Out of $10^{9}$ Muon Decays? {\em Phys. Lett. B} {\bf 1977}, {\em 67}, 421.
\bibitem{Fukugita:1986hr}
Fukugita, M.; Yanagida, T. Baryogenesis Without Grand Unification. {\em Phys. Lett. B} {\bf 1986}, {\em 174}, 45.
\bibitem{Coleman:1967ad}
Coleman, S. R.; Mandula, J. All possible symmetries of the S matrix. {\em Phys.\ Rev.} {\bf 1967}, {\em 159}, 1251.
\bibitem{susy-grav1}
Nath, P.; Arnowitt, R. L. Generalized Supergauge Symmetry As A New Framework For Unified Gauge Theories. {\em Phys. Lett. B}
{\bf 1975}, {\em 56}, 177.
\bibitem{susy-grav2}
Freedman, D. Z.; van Nieuwenhuizen, P.; Ferrara, S. Progress Toward A Theory Of Supergravity.
{\em Phys. Rev. D} {\bf 1976}, {\em 13}, 3214.
\bibitem{susy-grav3}
Deser, S.; Zumino, B. Consistent Supergravity. {\em Phys. Lett. B} {\bf 1976}, {\em 62}, 335.
\bibitem{Green:1987sp}
Green, M. B.; Schwarz, J. H.; Witten E. {\em Superstring Theory}; Cambridge Univ. Press, Cambridge, UK, 1987.
\bibitem{delAguila:1985hkb}
del Aguila, F.; Blair, G. A.; Daniel M.; Ross G. G. Superstring Inspired Models. {\em Nucl. Phys. B} {\bf 1986}, {\em 272}, 413.
\bibitem{Barbieri:1982eh}
Barbieri, R.; Ferrara, S.; Savoy, C. A. Gauge Models with Spontaneously Broken Local Supersymmetry. {\em Phys. Lett. B} {\bf 1982}, {\em 119}, 343.
\bibitem{Nilles:1982dy}
Nilles, H. P.; Srednicki M.; Wyler, D. Weak Interaction Breakdown Induced by Supergravity. {\em Phys. Lett. B} {\bf 1983}, {\em 120}, 346.
\bibitem{Hall:1983iz}
Hall, L. J.; Lykken J. D.; Weinberg, S. Supergravity as the Messenger of Supersymmetry Breaking. {\em Phys. Rev. D} {\bf 1983}, {\em 27}, 2359.
\bibitem{Soni:1983rm}
Soni, S. K.; Weldon, H. A. Analysis of the Supersymmetry Breaking Induced by N=1 Supergravity Theories. {\em Phys. Lett. B} {\bf 1983}, {\em 126}, 215.
\bibitem{Nilles:1990zd}
Nilles, H. P. Gaugino Condensation and Supersymmetry Breakdown. {\em Int. J. Mod. Phys. A} {\bf 1990}, {\em 5}, 4199.
\bibitem{Ellis:1990wk}
Ellis, J. R.; Kelley S.; Nanopoulos D. V. Probing the desert using gauge coupling unification. {\em Phys. Lett. B} {\bf 1991}, {\em 260}, 131.
\bibitem{Langacker:1991an}
Langacker, P.; Luo, M. X. Implications of precision electroweak experiments for $M_t$, $\rho_{0}$, $\sin^2\theta_W$ and grand unification.
{\em Phys. Rev. D} {\bf 1991}, {\em 44}, 817.
\bibitem{Amaldi:1991cn}
Amaldi, U.; de Boer, W.; Furstenau, H. Comparison of grand unified theories with electroweak and strong coupling constants measured at LEP.
{\em Phys. Lett. B} {\bf 1991}, {\em 260}, 447.
\bibitem{Anselmo:1991uu}
Anselmo, F.; Cifarelli, L.; Peterman, A.; Zichichi, A. The Effective experimental constraints on $M_(susy)$ and $M_(gut)$.
{\em Nuovo Cim. A} {\bf 1991}, {\em 104}, 1817.
\bibitem{Salam:1974jj}
Salam, A.; Strathdee, J. A. On Superfields and Fermi-Bose Symmetry. {\em Phys. Rev. D} {\bf 1975}, {\em 11}, 1521.
\bibitem{Grisaru:1979wc}
Grisaru, M. T.; Siegel W.; Rocek, M. Improved Methods for Supergraphs. {\em Nucl. Phys. B} {\bf 1979}, {\em 159}, 429.
\bibitem{Ellwanger:2009dp}
Ellwanger, U.; Hugonie C.; Teixeira, A. M. The Next-to-Minimal Supersymmetric Standard Model. {\em Phys. Rept.} {\bf 2010}, {\em 496}, 1.
\bibitem{Zeldovich:1974uw}
Zeldovich, Y. B.; Kobzarev, I. Y.; Okun, L. B. Cosmological Consequences of the Spontaneous Breakdown of Discrete Symmetry.
{\em Sov. Phys. JETP} {\bf 1974}, {\em 40}, 1.
\bibitem{Vilenkin:1984ib}
Vilenkin, A. Cosmic Strings and Domain Walls. {\em Phys. Rept.} {\bf 1985}, {\em 121}, 263.
\bibitem{Panagiotakopoulos:1998yw}
Panagiotakopoulos, C; Tamvakis, K. Stabilized NMSSM without domain walls. {\em Phys. Lett. B} {\bf 1999}, {\em 446}, 224.
\bibitem{Panagiotakopoulos:1999ah}
Panagiotakopoulos, C.; Tamvakis, K. New minimal extension of MSSM. {\em Phys. Lett. B} {\bf 1999}, {\em 469}, 145.
\bibitem{Hewett:1988xc}
Hewett, J. L.; Rizzo, T. G. Low-Energy Phenomenology of Superstring Inspired E(6) Models. {\em Phys. Rept.} {\bf 1989}, {\em 183}, 193.
\bibitem{01}
Binetruy, P.; Dawson, S.; Hinchliffe, I.; Sher, M. Phenomenologically Viable Models From Superstrings? {\em Nucl. Phys. B} {\bf 1986}, {\em 273}, 501.
\bibitem{02}
Ellis, J. R.; Enqvist, K.; Nanopoulos, D. V.; Zwirner, F. Observables In Low-Energy Superstring Models. {\em Mod. Phys. Lett. A} {\bf 1986}, {\em 1}, 57.
\bibitem{03}
Ellis, J. R.; Enqvist, K.; Nanopoulos, D. V.; Zwirner, F. Aspects Of The Superunification Of Strong, Electroweak And Gravitational
Interactions. {\em Nucl. Phys. B} {\bf 1986}, {\em 276}, 14.
\bibitem{04}
Ibanez, L. E.; Mas, J. Low-Energy Supergravity And Superstring Inspired Models. {\em Nucl. Phys. B} {\bf 1987}, {\em 286}, 107.
\bibitem{05}
Gunion, J. F.; Roszkowski, L.; Haber, H. E. Z-Prime Mass Limits, Masses And Couplings Of Higgs Bosons, And Z-Prime Decays
In An E(6) Superstring Based Model. {\em Phys. Lett. B} {\bf 1987}, {\em 189}, 409.
\bibitem{06}
Haber, H. E.; Sher, M. Higgs Mass Bound In E(6) Based Supersymmetric Theories. {\em Phys. Rev. D} {\bf 1987}, {\em 35}, 2206.
\bibitem{07}
Ellis, J. R.; Nanopoulos, D. V.; Petcov, S. T.; Zwirner, F. Gauginos And Higgs Particles In Superstring Models. {\em Nucl. Phys. B}
{\bf 1987}, {\em 283}, 93.
\bibitem{08}
Drees, M. Comment On 'Higgs Boson Mass Bound In E(6) Based Supersymmetric Theories. {\em Phys. Rev. D} {\bf 1987}, {\em 35}, 2910.
\bibitem{09}
Baer, H.; Dicus, D.; Drees, M.; Tata, X. Higgs Boson Signals In Superstring Inspired Models At Hadron Supercolliders.
{\em Phys. Rev. D} {\bf 1987}, {\em 36}, 1363.
\bibitem{010}
Gunion, J. F.; Roszkowski, L.; Haber, H. E. Production And Detection Of The Higgs Bosons Of The Simplest E(6) Based Gauge
Theory, {\em Phys. Rev. D} {\bf 1988}, {\em 38}, 105.
\bibitem{Langacker:1998tc}
Langacker, P.; Wang, J. U(1)' symmetry breaking in supersymmetric E(6) models, {\em Phys. Rev. D} {\bf 1998}, {\em 58}, 115010.
\bibitem{Cvetic:1995rj}
Cvetic, M.; Langacker, P. Implications of Abelian extended gauge structures from string models. {\em Phys. Rev. D} {\bf 1996}, {\em 54}, 3570.
\bibitem{Cvetic:1996mf}
Cvetic, M.; Langacker, P. New gauge bosons from string models, {\em Mod. Phys. Lett. A} {\bf 1996}, {\em 11}, 1247.
\bibitem{Cvetic:1997ky}
Cvetic, M.; Demir, D. A.; Espinosa, J. R.; Everett L. L.; Langacker, P. Electroweak breaking and the mu problem in supergravity models
with an additional U(1). {\em Phys. Rev. D} {\bf 1997}, {\em 56}, 2861.
\bibitem{Suematsu:1994qm}
Suematsu, D.; Yamagishi, Y. Radiative symmetry breaking in a supersymmetric model with an extra U(1). {\em Int. J. Mod. Phys. A}
{\bf 1995}, {\em 10}, 4521.
\bibitem{Keith:1997zb}
Keith, E.; Ma, E. Generic consequences of a supersymmetric U(1) gauge factor at the TeV scale. {\em Phys. Rev. D} {\bf 1997}, {\em 56}, 7155.
\bibitem{Daikoku:2000ep}
Daikoku, Y.; Suematsu, D. Mass bound of the lightest neutral Higgs scalar in the extra U(1) models. {\em Phys. Rev. D} {\bf 2000}, {\em 62}, 095006.
\bibitem{Kang:2004ix}
Kang, J. H.; Langacker, P.; Li, T. J. Neutrino masses in supersymmetric $SU(3)_{C} \times SU(2)_{L} \times U(1)_{Y} \times U(1)'$ models.
{\em Phys. Rev. D} {\bf 2005}, {\em 71}, 015012.
\bibitem{Ma:1995xk}
Ma, E. Neutrino masses in an extended gauge model with E(6) particle content. {\em Phys. Lett. B} {\bf 1996}, {\em 380}, 286.
\bibitem{Stech:2008wd}
Stech B., Tavartkiladze, Z. Generation Symmetry and $E_6$ Unification. {\em Phys. Rev. D} {\bf 2008}, {\em 77}, 076009.
\bibitem{Hambye:2000bn}
Hambye, T.; Ma, E.; Raidal, M.; Sarkar, U. Allowable low-energy E(6) subgroups from leptogenesis. {\em Phys. Lett. B} {\bf 2001}, {\em 512}, 373.
\bibitem{King:2008qb}
King, S. F.; Luo, R.; Miller, D. J.; Nevzorov, R. Leptogenesis in the Exceptional Supersymmetric Standard Model: flavour
dependent lepton asymmetries. {\em JHEP} {\bf 2008}, {\em 0812}, 042.
\bibitem{Nevzorov:2017gir}
Nevzorov, R. Leptogenesis as an origin of hot dark matter and baryon asymmetry in the $E_6$ inspired SUSY models.
{\em Phys. Lett. B} {\bf 2018}, {\em 779}, 223.
\bibitem{Nevzorov:2018leq}
Nevzorov, R. $E_6$ inspired SUSY models with custodial symmetry. {\em Int. J. Mod. Phys. A} {\bf 2018}, {\em 33}, 1844007.
\bibitem{Ma:2000jf}
Ma, E; Raidal, M. Three active and two sterile neutrinos in an E(6) model of diquark baryogenesis. {\em J. Phys. G} {\bf 2002}, {\em 28}, 95.
\bibitem{Kang:2004pp}
Kang, J.; Langacker, P.; Li T. J.; Liu, T. Electroweak baryogenesis in a supersymmetric U(1)-prime model. {\em Phys. Rev. Lett.} {\bf 2005},
{\em 94}, 061801.
\bibitem{Accomando:2010fz}
Accomando, E.; Belyaev, A.; Fedeli, L.; King, S. F.; Shepherd-Themistocleous, C. Z' physics with early LHC data.
{\em Phys. Rev. D} {\bf 2011}, {\em 83}, 075012.
\bibitem{Kang:2007ib}
Kang, J.; Langacker, P.; Nelson, B. D. Theory and Phenomenology of Exotic Isosinglet Quarks and Squarks.
{\em Phys. Rev. D} {\bf 2008}, {\em 77}, 035003.
\bibitem{g-2-1}
Grifols, J. A.; Sola, J.; Mendez, A. Contribution to the muon anomaly from superstring inspired models. {\em Phys. Rev. Lett.} {\bf 1986}, {\em 57}, 2348.
\bibitem{g-2-2}
Morris, D. A. Potentially large contributions to the muon anomalous magnetic moment from weak isosinglet squarks in E(6) superstring models.
{\em Phys. Rev. D} {\bf 1988}, {\em 37}, 2012.
\bibitem{Suematsu:1997tv}
Suematsu, D. Effect on the electron EDM due to abelian gauginos in SUSY extra U(1) models. {\em Mod. Phys. Lett. A} {\bf 1997}, {\em 12}, 1709.
\bibitem{GutierrezRodriguez:2006hb}
Gutierrez-Rodriguez, A.; Hernandez-Ruiz M. A.; Perez, M. A. Limits on the Electromagnetic and Weak Dipole Moments of the Tau-Lepton in
E(6) Superstring Models. {\em Int. J. Mod. Phys. A} {\bf 2007}, {\em 22}, 3493.
\bibitem{Suematsu:1997qt}
Suematsu, D. $\mu \to e \gamma$ in supersymmetric multi U(1) models with an abelian gaugino mixing.
{\em Phys. Lett. B} {\bf 1998}, {\em 416}, 108.
\bibitem{Ham:2008fx}
Ham, S. W.; Im, J. O.; Yoo E. J.; Oh, S. K. Higgs bosons of a supersymmetric $E_6$ model at the Large Hadron Collider.
{\em JHEP} {\bf 2008}, {\em 0812}, 017.
\bibitem{Suematsu:1997au}
Suematsu, D. Neutralino decay in the mu problem solvable extra U(1) models. {\em Phys. Rev. D} {\bf 1998}, {\em 57}, 1738.
\bibitem{Keith:1996fv}
Keith, E.; Ma, E. Efficacious Extra U(1) Factor for the Supersymmetric Standard Model. {\em Phys. Rev. D} {\bf 1996}, {\em 54}, 3587.
\bibitem{n1}
Hesselbach, S.; Franke, F.; Fraas, H. Neutralinos in E(6) inspired supersymmetric U(1)' models.
{\em Eur. Phys. J. C} {\bf 2002}, {\em 23}, 149.
\bibitem{n2}
Barger, V.; Langacker, P.; Lee, H. S. Lightest neutralino in extensions of the MSSM. {\em Phys. Lett. B} {\bf 2005}, {\em 630}, 85.
\bibitem{n3}
Choi, S. Y.; Haber, H. E.; Kalinowski, J.; Zerwas, P. M. The neutralino sector in the U(1)-extended supersymmetric standard model.
{\em Nucl. Phys. B} {\bf 2007}, {\em 778}, 85.
\bibitem{n4}
Barger, V.; Langacker, P.; Lewis, I.; McCaskey, M.; Shaughnessy G.; Yencho, B. Recoil detection of the lightest neutralino in MSSM
singlet extensions. {\em Phys. Rev. D} {\bf 2007}, {\em 75}, 115002.
\bibitem{Gherghetta:1996yr}
Gherghetta, T.; Kaeding, T. A.; Kane, G. L. Supersymmetric contributions to the decay of an extra $Z$ boson. {\em Phys. Rev. D} {\bf 1998}, {\em 57}, 3178.
\bibitem{E6neutralino-higgs}
Barger, V.; Langacker, P.; Shaughnessy, G. TeV physics and the Planck scale. {\em New J. Phys.} {\bf 2007}, {\em 9}, 333.
\bibitem{King:2005jy}
King, S. F.; Moretti, S.; Nevzorov, R. Theory and phenomenology of an exceptional supersymmetric standard model.
{\em Phys. Rev. D} {\bf 2006}, {\em 73}, 035009.
\bibitem{King:2005my}
King, S. F.; Moretti, S.; Nevzorov, R. Exceptional supersymmetric standard model. {\em Phys. Lett. B} {\bf 2006}, {\em 634}, 278.
\bibitem{E6-higgs}
Barger, V.; Langacker, P.; Lee, H. S.; Shaughnessy, G. Higgs Sector in Extensions of the MSSM. {\em Phys. Rev. D} {\bf 2006}, {\em 73}, 115010.
\bibitem{Nevzorov:2012hs}
Nevzorov, R. $E_6$ inspired supersymmetric models with exact custodial symmetry. {\em Phys. Rev. D} {\bf 2013}, {\em 87}, 015029.
\bibitem{Howl:2007hq}
Howl, R.; King, S. F. Planck Scale Unification in a Supersymmetric Standard Model. {\em Phys. Lett. B} {\bf 2007}, {\em 652}, 331.
\bibitem{Howl:2007zi}
Howl, R.; King, S. F. Minimal $E_6$ Supersymmetric Standard Model. {\em JHEP} {\bf 2008}, {\em 0801}, 030.
\bibitem{Howl:2008xz}
Howl, R.; King, S. F. Exceptional Supersymmetric Standard Models with non-Abelian Discrete Family Symmetry. {\em JHEP} {\bf 2008}, {\em 0805}, 008.
\bibitem{Howl:2009ds}
Howl, R.; King, S. F. Solving the Flavour Problem in Supersymmetric Standard Models with Three Higgs Families. {\em Phys. Lett. B} {\bf 2010},
{\em 687}, 355.
\bibitem{Athron:2010zz}
Athron, P.; Hall, J. P.; Howl, R.; King, S. F.; Miller, D. J.; Moretti, S.; Nevzorov, R. Aspects of the Exceptional Supersymmetric Standard
Model. {\em Nucl. Phys. Proc. Suppl.} {\bf 2010}, {\em 200-202}, 120.
\bibitem{Hall:2011zq}
Hall, J. P.; King, S. F. Bino Dark Matter and Big Bang Nucleosynthesis in the Constrained E$_6$SSM with Massless Inert Singlinos. {\em JHEP} {\bf 2011},
{\em 1106}, 006.
\bibitem{Callaghan:2012rv}
Callaghan, J. C.; King, S. F. $E_6$ Models from F-theory. {\em JHEP} {\bf 2013}, {\em 1304}, 034.
\bibitem{Callaghan:2013kaa}
Callaghan, J. C.; King, S. F.; Leontaris, G. K. Gauge coupling unification in $E_6$ F-theory GUTs with matter and bulk exotics from flux breaking.
{\em JHEP} {\bf 2013}, {\em 1312}, 037.
\bibitem{Athron:2014pua}
Athron, P.; M{\"u}hlleitner, M.; Nevzorov, R.; Williams, A. G. Non-Standard Higgs Decays in $U(1)$ Extensions of the MSSM.
{\em JHEP} {\bf 2015}, {\em 1501}, 153.
\bibitem{King:2016wep}
King, S. F.; Nevzorov, R. 750 GeV Diphoton Resonance from Singlets in an Exceptional Supersymmetric Standard Model. {\em JHEP} {\bf 2016},
{\em 1603}, 139.
\bibitem{Hall:2009aj}
Hall, J. P.; King, S. F. Neutralino Dark Matter with Inert Higgsinos and Singlinos. {\em JHEP} {\bf 2009}, {\em 0908}, 088.
\bibitem{King:2007uj}
King, S. F.; Moretti, S.; Nevzorov, R. Gauge coupling unification in the exceptional supersymmetric standard model. {\em Phys. Lett. B} {\bf 2007},
{\em 650}, 57.
\bibitem{Nevzorov:2013ixa}
Nevzorov, R. Quasifixed point scenarios and the Higgs mass in the $E_6$ inspired supersymmetric models. {\em Phys. Rev. D} {\bf 2014}, {\em 89}, 055010.
\bibitem{Nevzorov:2015iya}
Nevzorov, R. LHC Signatures and Cosmological Implications of the E$_6$ Inspired SUSY Models. {\em PoS EPS} {\bf 2015}, {\em -HEP2015}, 381.
\bibitem{Nevzorov:2001vj}
Nevzorov, R.; Trusov, M. A. Infrared quasifixed solutions in the NMSSM. {\em Phys. Atom. Nucl.} {\bf 2001}, {\em 64}, 1299.
\bibitem{Nevzorov:2002ub}
Nevzorov, R.; Trusov, M. A. Quasifixed point scenario in the modified NMSSM. {\em Phys. Atom. Nucl.} {\bf 2002}, {\em 65}, 335.
\bibitem{King:2006vu}
King, S. F.; Moretti, S.; Nevzorov, R. Spectrum of Higgs particles in the ESSM. hep-ph/0601269.
\bibitem{King:2006rh}
King, S. F.; Moretti, S.; Nevzorov, R. E$_6$SSM. {\em  AIP Conf. Proc.} {\bf 2007}, {\em 881}, 138.
\bibitem{Athron:2011wu}
Athron, P.; King, S. F.; Miller, D. J.; Moretti, S.; Nevzorov, R. LHC Signatures of the Constrained Exceptional Supersymmetric Standard Model.
{\em Phys. Rev. D} {\bf 2011}, {\em 84}, 055006.
\bibitem{Belyaev:2012si}
Belyaev, A.; Hall, J. P.; King, S. F.; Svantesson, P. Novel gluino cascade decays in $E_6$ inspired models. {\em Phys. Rev. D}
{\bf 2012}, {\em 86}, 031702.
\bibitem{Belyaev:2012jz}
Belyaev, A.; Hall, J. P.; King, S. F.; Svantesson, P. Discovering $E_6$ supersymmetric models in gluino cascade decays at the LHC. {\em Phys. Rev. D}
{\bf 2013}, {\em 87}, 035019.
\bibitem{Hall:2010ix}
Hall, J. P.; King, S. F.; Nevzorov, R.; Pakvasa, S.; Sher, M. Novel Higgs Decays and Dark Matter in the E$_6$SSM. {\em Phys. Rev. D} {\bf 2011}, {\em 83}, 075013.
\bibitem{Hall:2010ny}
Hall, J. P.; King, S. F.; Nevzorov, R.; Pakvasa, S.; Sher, M. Nonstandard Higgs decays in the E$_6$SSM. {\em PoS } {\bf 2010}, {\em QFTHEP2010}, 069.
\bibitem{Hall:2011au}
Hall, J. P.; King, S. F.; Nevzorov, R.; Pakvasa, S.; Sher, M. Nonstandard Higgs Decays and Dark Matter in the E$_6$SSM. arXiv:1109.4972.
\bibitem{Hall:2013bua}
Hall, J. P.; King, S. F.; Nevzorov, R.; Pakvasa, S.; Sher, M. Dark matter and nonstandard Higgs decays in the exceptional supersymmetric
standard model. {\em AIP Conf. Proc.} {\bf 2013}, {\em 1560}, 303.
\bibitem{Nevzorov:2013tta}
Nevzorov, R.; Pakvasa, S. Exotic Higgs decays in the $E_6$ inspired SUSY models. {\em Phys. Lett. B} {\bf 2014}, {\em 728}, 210.
\bibitem{Nevzorov:2014sha}
Nevzorov, R.; Pakvasa, S. Nonstandard Higgs decays in the $E_6$ inspired SUSY models. {\em Nucl. Part. Phys. Proc.} {\bf 2016}, {\em 273-275}, 690.
\bibitem{Athron:2016usd}
Athron, P.; Muhlleitner, M.; Nevzorov, R.; Williams, A. G. Exotic Higgs decays in U(1) extensions of the MSSM. arXiv:1602.04453.
\bibitem{Athron:2008np}
Athron, P.; King, S. F.; Miller, D. J.; Moretti, S.; Nevzorov, R. The Constrained E$_6$SSM. arXiv:0810.0617.
\bibitem{Athron:2009ue}
Athron, P.; King, S. F.; Miller, D. J.; Moretti, S.; Nevzorov, R. Predictions of the Constrained Exceptional Supersymmetric Standard Model.
{\em Phys. Lett. B} {\bf 2009}, {\em 681}, 448.
\bibitem{Athron:2009bs}
Athron, P.; King, S. F.; Miller, D. J.; Moretti, S.; Nevzorov, R. The Constrained Exceptional Supersymmetric Standard Model.
{\em Phys. Rev. D} {\bf 2009}, {\em 80}, 035009.
\bibitem{Athron:2012sq}
Athron, P.; King, S. F.; Miller, D. J.; Moretti, S.; Nevzorov, R. Constrained Exceptional Supersymmetric Standard Model with a Higgs Near 125 GeV.
{\em Phys. Rev. D} {\bf 2012}, {\em 86}, 095003.
\bibitem{Athron:2015vxg}
Athron, P.; Harries, D.   ; Nevzorov, R.; Williams, A. G. $E_6$ Inspired SUSY benchmarks, dark matter relic density and a 125 GeV Higgs.
{\em Phys. Lett. B} {\bf 2016}, {\em 760}, 19.
\bibitem{Athron:2016gor}
Athron, P.; Harries, D.   ; Nevzorov, R.; Williams, A. G. Dark matter in a constrained $E_{6}$ inspired SUSY model.
{\em JHEP} {\bf 2016}, {\em 1612}, 128.
\bibitem{Athron:2013ipa}
Athron, P.; Binjonaid, M.; King, S. F. Fine Tuning in the Constrained Exceptional Supersymmetric Standard Model. {\em Phys. Rev. D}
{\bf 2013}, {\em 87}, 115023.
\bibitem{Athron:2015tsa}
Athron, P.; Harries, D.; Williams, A. G. $Z^\prime$ mass limits and the naturalness of supersymmetry. {\em Phys. Rev. D}
{\bf 2015}, {\em 91}, 115024.
\bibitem{Athron:2012pw}
Athron, P.; St{\"o}ckinger, D.; Voigt, A. Threshold Corrections in the Exceptional Supersymmetric Standard Model.
{\em Phys. Rev. D} {\bf 2012}, {\em 86}, 095012.
\bibitem{Sperling:2013eva}
Sperling, M.; St{\"o}ckinger, D.; Voigt, A. Renormalization of vacuum expectation values in spontaneously broken gauge theories.
{\em JHEP} {\bf 2013}, {\em 1307}, 132.
\bibitem{Sperling:2013xqa}
Sperling, M.; St{\"o}ckinger, D.; Voigt, A. Renormalization of vacuum expectation values in spontaneously broken gauge theories: Two-loop
results. {\em JHEP} {\bf 2014}, {\em 1401}, 068.
\bibitem{431}
S.~Wolfram,  Phys.\ Lett.\ B {\bf 82} (1979) 65;
C.~B.~Dover, T.~K.~Gaisser, G.~Steigman, Phys.\ Rev.\ Lett. {\bf 42} (1979) 1117.
\bibitem{Wolfram:1978}
Wolfram, S. Abundances of Stable Particles Produced in the Early Universe. {\em Phys. Lett. B} {\bf 1979}, {\em 82} 65.
\bibitem{Dover:1979sn}
Dover, C. B.; Gaisser, T. K.; Steigman, G. Cosmological Constraints on New Stable Hadrons. {\em Phys. Rev. Lett.} {\bf 1979}, {\em 42}, 1117.
\bibitem{Rich:1987jd}
Rich, J.; Lloyd Owen, D.; Spiro, M. Experimental particle physics without accelerators. {\em Phys. Rept.} {\bf 1987}, {\em 151}, 239.
\bibitem{Smith:1988ni}
Smith, P. F. Terrestrial Searches for New Stable Particles. {\em Contemp. Phys.} {\bf 1988}, {\em 29}, 159.
\bibitem{Hemmick:1989ns}
Hemmick, T. K. et al. A Search for Anomalously Heavy Isotopes of Low $Z$ Nuclei. {\em Phys. Rev. D} {\bf 1990}, {\em 41}, 2074.
\bibitem{Giudice:1988yz}
Giudice, G. F.; Masiero, A. A Natural Solution to the $\mu$ Problem in Supergravity Theories. {\em Phys. Lett. B} {\bf 1988}, {\em 206}, 480.
\bibitem{Casas:1992mk}
Casas, J. A.; Mu\~{n}oz, C. A Natural solution to the $\mu$ problem. {\em Phys. Lett. B} {\bf 1993}, {\em 306}, 288.
\bibitem{Hesselbach:2007te}
Hesselbach, S.; Miller, D. J.; Moortgat-Pick, G.; Nevzorov, R.; Trusov, M. Theoretical upper bound on the mass of the LSP in the MNSSM.
{\em Phys. Lett. B} {\bf 2008}, {\em 662}, 199.
\bibitem{Hesselbach:2007ta}
Hesselbach, S.; Miller, D. J.; Moortgat-Pick, G.; Nevzorov, R.; Trusov, M. The Lightest neutralino in the MNSSM. arXiv:0710.2550.
\bibitem{Hesselbach:2008vt}
Hesselbach, S.; Miller, D. J.; Moortgat-Pick, G.; Nevzorov, R.; Trusov, M. Lightest Neutralino Mass in the MNSSM. arXiv:0810.0511.
\bibitem{Frere:1996gb}
Frere, J. M.; Nevzorov, R.; Vysotsky, M. I. Stimulated neutrino conversion and bounds on neutrino magnetic moments.
{\em Phys. Lett. B} {\bf 1997}, {\em 394}, 127.
\bibitem{Holdom:1985ag}
Holdom, B. Two U(1)'s and Epsilon Charge Shifts. {\em Phys. Lett. B} {\bf 1986}, {\em 166}, 196.
\bibitem{Babu:1996vt}
Babu, K. S.; Kolda C. F.; March-Russell, J. Leptophobic U(1) and the R($b$) - R($c$) crisis. {\em Phys. Rev. D} {\bf 1996}, {\em 54}, 4635.
\bibitem{Babu:1997st}
Babu, K. S.; Kolda, C. F.; March-Russell, J. Implications of generalized $Z-Z'$ mixing {\em Phys. Rev. D} {\bf 1998}, {\em 57}, 6788.
\bibitem{Rizzo:1998ut}
Rizzo, T. G. Gauge kinetic mixing and leptophobic $Z^\prime$ in E(6) and SO(10). {\em Phys. Rev. D} {\bf 1998}, {\em 59}, 015020.
\bibitem{Suematsu:1998wm}
Suematsu, D. Vacuum structure of the $\mu$ problem solvable extra U(1) models. {\em Phys. Rev. D} {\bf 1999}, {\em 59}, 055017.
\bibitem{Martin:1993zk}
Martin, S. P.; Vaughn, M. T. Two loop renormalization group equations for soft supersymmetry breaking couplings.
{\em Phys. Rev. D} {\bf 1994}, {\em 50}, 2282.
\bibitem{Chankowski:1995dm}
Chankowski, P. H.; Pluciennik, Z.; Pokorski, S.; Vayonakis, C. E. Gauge coupling unification in GUT and string models.
{\em Phys. Lett. B} {\bf 1995}, {\em 358}, 264.
\bibitem{Antoniadis:1982vr}
Antoniadis, I.; Kounnas, C.; Tamvakis, K. Simple Treatment of Threshold Effects.
{\em Phys. Lett. B} {\bf 1982}, {\em 119}, 377.
\bibitem{Antoniadis:1982qw}
Antoniadis, I.; Kounnas, C.; Lacaze, R. Light Gluinos in Deep Inelastic Scattering.
{\em Nucl. Phys. B} {\bf 1983}, {\em 211}, 216.
\bibitem{Carena:1993ag}
Carena, M.; Pokorski, S.; Wagner, C. E. M. On the unification of couplings in the minimal supersymmetric Standard Model.
{\em Nucl. Phys. B} {\bf 1993}, {\em 406}, 59.
\bibitem{Bagger:1995bw}
Bagger, J.; Matchev K. T.; Pierce, D. Precision corrections to supersymmetric unification.
{\em Phys. Lett. B} {\bf 1995}, {\em 348}, 443.
\bibitem{Langacker:1995fk}
Langacker, P.; Polonsky, N. The Strong coupling, unification, and recent data.
{\em Phys. Rev. D} {\bf 1995}, {\em 52}, 3081.
\bibitem{Langacker:1992rq}
Langacker, P.; Polonsky, N. Uncertainties in coupling constant unification.
{\em Phys. Rev. D} {\bf 1993}, {\em 47}, 4028.
\bibitem{gc-1}
Ross, G. G.; Roberts, R. G. Minimal supersymmetric unification predictions.
{\em Nucl. Phys. B} {\bf 1992}, {\em 377}, 571.
\bibitem{gc-2}
Barger, V. D.; Berger, M. S.; Ohmann, P. Supersymmetric grand unified theories: Two loop evolution of gauge and Yukawa couplings.
{\em Phys. Rev. D} {\bf 1993}, {\em 47}, 1093.
\bibitem{gc-3}
Langacker, P.; Polonsky, N. The Bottom mass prediction in supersymmetric grand unification: Uncertainties and constraints.
{\em Phys. Rev. D} {\bf 1994}, {\em 49}, 1454.
\bibitem{deBoer:2003xm}
de Boer W.; Sander, C. Global electroweak fits and gauge coupling unification.
{\em Phys. Lett. B} {\bf 2004}, {\em 585}, 276.
\bibitem{deBoer:2005bd}
de Boer, W.; Sander, C.; Zhukov, V.; Gladyshev, A. V.; Kazakov, D. I. The Supersymmetric interpretation of the EGRET excess
of diffuse galactic gamma rays. {\em Phys. Lett. B} {\bf 2006}, {\em 636}, 13.
\bibitem{Kovalenko:1998dc}
Kovalenko, P. A.; Nevzorov R. B.; Ter-Martirosian, K. A. Masses of Higgs bosons in supersymmetric theories.
{\em Phys. Atom. Nucl.} {\bf 1998}, {\em 61}, 812.
\bibitem{Nevzorov:2000uv}
Nevzorov, R. B.; Trusov, M. A. Particle spectrum in the modified NMSSM in the strong Yukawa coupling limit.
{\em J. Exp. Theor. Phys.} {\bf 2000}, {\em 91}, 1079.
\bibitem{Nevzorov:2001um}
Nevzorov, R. B.; Ter-Martirosyan, K. A.; Trusov, M. A. Higgs bosons in the simplest SUSY models.
{\em Phys. Atom. Nucl.} {\bf 2002}, {\em 65}, 285.
\bibitem{Miller:2003ay}
Miller, D. J.; Nevzorov, R.; Zerwas, P. M. The Higgs sector of the next-to-minimal supersymmetric standard model.
{\em Nucl. Phys. B} {\bf 2004}, {\em 681}, 3
\bibitem{Nevzorov:2004ge}
Nevzorov, R.; Miller, D. J. Approximate solutions for the Higgs masses and couplings in the NMSSM. hep-ph/0411275.
\bibitem{Miller:2005qua}
Miller, D. J.; Moretti, S.; Nevzorov, R. Higgs bosons in the NMSSM with exact and slightly broken PQ-symmetry. hep-ph/0501139.
\bibitem{Miller:2003hm}
Miller, D. J.; Nevzorov, R. The Peccei-Quinn axion in the next-to-minimal supersymmetric standard model. hep-ph/0309143.
\bibitem{King:2014xwa}
King, S. F.; Muhlleitner, M.; Nevzorov R.; Walz, K. Discovery Prospects for NMSSM Higgs Bosons at the High-Energy Large Hadron Collider.
{\em Phys. Rev. D} {\bf 2014}, {\em 90}, 095014.
\bibitem{Durand:1988rg}
Durand, L.; Lopez, J. L. Upper Bounds on Higgs and Top Quark Masses in the Flipped SU(5) x U(1) Superstring Model.
{\em Phys. Lett. B} {\bf 1989}, {\em 217}, 463.
\bibitem{Drees:1988fc}
Drees, M. Supersymmetric Models with Extended Higgs Sector. {\em Int. J. Mod. Phys. A} {\bf 1989}, {\em 4}, 3635.
\bibitem{Flores:1982pr}
Flores, R. A.; Sher, M. Higgs Masses in the Standard, Multi-Higgs and Supersymmetric Models. {\em Annals Phys.} {\bf 1983}, {\em 148}, 95.
\bibitem{Inoue:1982ej}
Inoue, K.; Kakuto, A.; Komatsu, H.; Takeshita, S. Low-Energy Parameters and Particle Masses in a Supersymmetric Grand Unified Model.
{\em Prog. Theor. Phys.} {\bf 1982}, {\em 67}, 1889.
\bibitem{Djouadi:2005gj}
Djouadi, A. The Anatomy of electro-weak symmetry breaking. II. The Higgs bosons in the minimal supersymmetric model.
{\em Phys. Rept.} {\bf 2008}, {\em 459}, 1.
\bibitem{Tanabashi:2018oca}
Tanabashi, M. {\it et al.} [Particle Data Group] Review of Particle Physics.
{\em Phys. Rev. D} {\bf 2018}, {\em 98}, 030001.
\bibitem{mtMS-1}
Chetyrkin, K. G.; Steinhauser, M. Short distance mass of a heavy quark at order $\alpha_s^3$.
{\em Phys. Rev. Lett.} {\bf 1999}, {\em 83}, 4001.
\bibitem{mtMS-2}
Chetyrkin, K. G.; Steinhauser, M. The relation between the MS-bar and the on-shell quark mass at order $\alpha_s^3$.
{\em Nucl. Phys. B} {\bf 2000}, {\em 573}, 617.
\bibitem{Carena:1995wu}
Carena, M.; Quiros M.; Wagner, C. E. M. Effective potential methods and the Higgs mass spectrum in the MSSM.
{\em Nucl. Phys. B} {\bf 1996}, {\em 461}, 407.
\bibitem{Ellwanger:1999ji}
Ellwanger, U.; Hugonie, C. Masses and couplings of the lightest Higgs bosons in the (M+1)SSM.
{\em Eur. Phys. J. C} {\bf 2002}, {\em 25}, 297.
\bibitem{suspect}
Djouadi, A.; Kneur, J. L.; Moultaka, G. SuSpect: A Fortran code for the supersymmetric and Higgs particle spectrum in the MSSM.
{\em Comput. Phys. Commun.} {\bf 2007}, {\em 176}, 426.
\bibitem{feynhiggs-1}
Heinemeyer, S.; Hollik, W.; Weiglein, G. FeynHiggs: A Program for the calculation of the masses of the neutral CP even Higgs bosons in the MSSM.
{\em Comput. Phys. Commun.} {\bf 2000}, {\em 124}, 76.
\bibitem{feynhiggs-2}
Heinemeyer, S.; Hollik, W.; Weiglein, G. The Masses of the neutral CP - even Higgs bosons in the MSSM: Accurate analysis at the two loop level.
{\em Eur. Phys. J. C} {\bf 1999}, {\em 9}, 343.
\bibitem{feynhiggs-3}
Degrassi, G.; Heinemeyer, S.; Hollik, W.; Slavich, P.; Weiglein, G. Towards high precision predictions for the MSSM Higgs sector.
{\em Eur. Phys. J. C} {\bf 2003}, {\em 28}, 133.
\bibitem{feynhiggs-4}
Frank, M.; Hahn, T.; Heinemeyer, S.; Hollik, W.; Rzehak, H.; Weiglein, G.
The Higgs Boson Masses and Mixings of the Complex MSSM in the Feynman-Diagrammatic Approach.
{\em JHEP} {\bf 2007}, {\em 0702}, 047.
\bibitem{King:2012is}
King, S. F.; Muhlleitner, M.; Nevzorov, R. NMSSM Higgs Benchmarks Near 125 GeV.
{\em Nucl. Phys. B} {\bf 2012}, {\em 860}, 207.
\bibitem{King:2012wg}
King, S. F.; Merle, A. Warm Dark Matter from keVins. {\em JCAP} {\bf 2012}, {\em 1208}, 016.
\bibitem{Aaboud:2017buh}
Aaboud, M. {\it et al.} [ATLAS Collaboration]. Search for new high-mass phenomena in the dilepton final state using
$36\,\mbox{fb}^{-1}$ of proton-proton collision data at $ \sqrt{s}=13 $ TeV with the ATLAS detector.
{\em JHEP} {\bf 2017}, {\em 1710}, 182.
\bibitem{Sirunyan:2018exx}
Sirunyan, A. M. {\it et al.} [CMS Collaboration]. Search for high-mass resonances in dilepton final states in proton-proton
collisions at $\sqrt{s}=$ 13 TeV. {\em JHEP} {\bf 2018}, {\em 1806}, 120.
\bibitem{Accomando:2010fz}
Accomando, E.; Belyaev, A.; Fedeli, L.; King, S. F.; Shepherd-Themistocleous, C. Z' physics with early LHC data.
{\em Phys. Rev. D} {\bf 2011}, {\em 83}, 075012.
\bibitem{Gherghetta:1996yr}
Gherghetta, T.; Kaeding, T. A.; Kane, G. L. Supersymmetric contributions to the decay of an extra $Z$ boson.
{\em Phys. Rev. D} {\bf 1998}, {\em 57}, 3178.
\bibitem{Sirunyan:2018btu}
Sirunyan, A. M. {\it et al.} [CMS Collaboration]. Search for pair production of first-generation scalar leptoquarks at $\sqrt{s} =$13 TeV.
{\em Phys. Rev. D} {\bf 2019}, {\em 99} 052002.
\bibitem{Aaboud:2019jcc}
Aaboud, M. {\it et al.} [ATLAS Collaboration]. Searches for scalar leptoquarks and differential cross-section measurements in dilepton-dijet
events in proton-proton collisions at a centre-of-mass energy of $\sqrt{s}$ = 13 TeV with the ATLAS experiment.
{\em Eur. Phys. J. C} {\bf 2019}, {\em 79}, 733.
\bibitem{Sirunyan:2018ryt}
Sirunyan, A. M. {\it et al.} [CMS Collaboration]. Search for pair production of second-generation leptoquarks at $\sqrt{s}=$ 13 TeV.
{\em Phys. Rev. D} {\bf 2019}, {\em 99}, 032014.
\bibitem{Sirunyan:2018ruf}
Sirunyan, A. M. {\it et al.} [CMS Collaboration].
Search for leptoquarks coupled to third-generation quarks in proton-proton collisions at $\sqrt{s}=$ 13 TeV.
{\em Phys. Rev. Lett.} {\bf 2018}, {\em 121}, 241802.
\bibitem{Sirunyan:2018nkj}
Sirunyan, A. M. {\it et al.} [CMS Collaboration]. Search for third-generation scalar leptoquarks decaying to a top quark and
a $\tau$ lepton at $\sqrt{s}=$ 13 TeV. {\em Eur. Phys. J. C} {\bf 2018}, {\em 78}, 707.
\bibitem{Aaboud:2019bye}
Aaboud, M. {\it et al.} [ATLAS Collaboration]. Searches for third-generation scalar leptoquarks in $\sqrt{s}$ = 13 TeV pp collisions
with the ATLAS detector. {\em JHEP} {\bf 2019}, {\em 1906}, 144.
\bibitem{Sirunyan:2018xlo}
Sirunyan, A. M. {\it et al.} [CMS Collaboration]. Search for narrow and broad dijet resonances in proton-proton collisions
at $ \sqrt{s}=13 $ TeV and constraints on dark matter mediators and other new particles.
{\em JHEP} {\bf 2018}, {\em 1808}, 130.
\bibitem{Aaboud:2018jiw}
Aaboud, M. {\it et al.} [ATLAS Collaboration]. Search for electroweak production of supersymmetric particles in final states with two
or three leptons at $\sqrt{s}=13\,$TeV with the ATLAS detector.
{\em Eur. Phys. J. C} {\bf 2018}, {\em 78},  995.
\bibitem{Aaboud:2018sua}
Aaboud, M. {\it et al.} [ATLAS Collaboration]. Search for chargino-neutralino production using recursive jigsaw reconstruction in final states
with two or three charged leptons in proton-proton collisions at $\sqrt{s}=13$ TeV with the ATLAS detector. {\em Phys. Rev. D} {\bf 2018}, {\em 98}, 092012.
\bibitem{Sirunyan:2018ubx}
Sirunyan, A. M. {\it et al.} [CMS Collaboration]. Combined search for electroweak production of charginos and neutralinos in proton-proton
collisions at $\sqrt{s} =$ 13 TeV. {\em JHEP} {\bf 2018}, {\em 1803}, 160.
\bibitem{Kuzmin:1985mm}
Kuzmin, V. A.; Rubakov, V. A.; Shaposhnikov, M. E. On The Anomalous Electroweak Baryon Number Nonconservation in the Early
universe. {\em Phys. Lett. B} {\bf 1985}, {\em 155}, 36.
\bibitem{Rubakov:1996}
Rubakov, V. A.; Shaposhnikov, M. E. Electroweak baryon number non-conservation in the Early Universe and in high-energy collisions.
{\em Usp. Fiz. Nauk} {\bf 1996}, {\em 166}, 493.

\end{thebibliography}
\end{document}